# PKComb-BOIN12: A Pharmacokinetically Guided Bayesian Design for Dose Optimisation in Oncology Drug-Combination Trials

Jieqi Tu[1], Ruitao Lin[2], Ayon Mukherjee[3,*]

[1]Global Statistical Sciences, Eli Lilly and Company, Indianapolis, IN, USA
[2]Department of Biostatistics, The University of Texas MD Anderson Cancer Center, Houston, TX, USA
[3]Population Health Sciences Institute, Newcastle University, Newcastle upon Tyne, United Kingdom

[*]Correspondence: Ayon Mukherjee, Population Health Sciences Institute, Newcastle University, Newcastle upon Tyne, United Kingdom. Email: ayon.mukherjee@newcastle.ac.uk

**Funding:** National Institute for Health and Care Research, Grant/Award Number: NIHR301614

## ABSTRACT

Dose optimisation of two-agent combinations in early-phase oncology trials must jointly balance toxicity, efficacy, and pharmacologically meaningful drug exposure, yet existing model-assisted combination designs use only binary toxicity–efficacy outcomes and leave pharmacokinetic (PK) data unused for dosing decisions. We propose PKComb-BOIN12, a utility-based Bayesian optimal interval design embedding a continuous PK exposure metric into two-agent dose optimisation via a PK-guided escalation rule and a PK-sufficiency constraint on the final admissible set. Two contributions distinguish this design from a direct combination extension of existing single-agent PK-informed designs. First, PK-based elimination uses a coordinatewise order-ideal (down-set) removal rule, so that eliminating a combination for insufficient exposure is licensed exactly by the assumed monotonicity of exposure in each agent's dose, rather than an unjustified scalar summary of combined dose intensity. Second, the pre-specified target exposure is treated as transportable-but-verifiable: a hierarchical calibration mechanism updates the assumed monotherapy-derived target using accruing combination-arm PK data, with a formal bound on the resulting selection bias under transportability violations. A time-to-event extension, TITE-PKComb-BOIN12, accommodates late-onset outcomes. A factorial simulation study spanning realistic scenarios, including one calibrated to a published combination trial, decomposes the source of PKComb-BOIN12's gains over its non-PK predecessor and evaluates robustness to non-monotone exposure, PK-variance misspecification, and target-transportability violations. An R Shiny application supports design evaluation and real-time conduct.

**Keywords:** Bayesian adaptive design; Bayesian optimal interval design; dose optimisation; drug-combination trial; model-informed drug development; optimal biological dose combination; pharmacokinetics; Project Optimus

## 1 Introduction

Dose optimisation in early-phase oncology trials requires the simultaneous characterisation of safety, pharmacokinetic (PK) behaviour, and early evidence of therapeutic activity. Historically, dose escalation relied on rule-based algorithms such as the "3+3" design, valued for operational simplicity yet repeatedly shown to be statistically inefficient: it under-exposes participants relative to the true maximum tolerated dose (MTD), yields highly variable dose recommendations across trials, and ignores efficacy evidence entirely when selecting a dose [1, 2, 3]. Model-based designs, led by the continual reassessment method (CRM) [4] and its combination extensions [5, 6, 7], address these deficiencies through adaptive posterior updating but have seen limited routine uptake owing to computational demands and sensitivity to prior specification. Interval-based model-assisted designs bridge this gap: the Bayesian Optimal Interval (BOIN) design [8] derives decision boundaries from Bayesian optimality principles while retaining pre-tabulated escalation rules, and its phase I/II extension BOIN12 [9] augments these boundaries with a bivariate toxicity–efficacy utility framework that targets an optimal biological dose (OBD) rather than the MTD alone. The combination counterpart, Comb-BOIN12 [10], adapts this utility approach to two-agent trials via a rank-based desirability score (RDS), and TITE-BOIN12 [11] accommodates late-onset outcomes through an approximated likelihood, together establishing a practically deployable platform for combination dose optimisation.

A persistent limitation of this design family is that PK information routinely collected in early-phase trials plays no formal role in dosing decisions – neither in the interim allocation of successive cohorts nor in the determination of the dose recommended for later-phase evaluation. For classical cytotoxics this omission is defensible, since exposure and toxicity rise together and the MTD closely approximates the clinically optimal dose [12]. For targeted and immunotherapeutic agents, whose dose–response relationships are frequently nonlinear, saturable, or unimodal, an OBD defined jointly by adequate exposure and acceptable tolerability is more clinically relevant than the MTD [13, 14, 15]. Regulatory frameworks reinforce this view: the ICH E4 guideline [16] and the U.S. Food and Drug Administration's (FDA) Project Optimus initiative [17] both call for exposure–response evidence to support dose selection, and the model-informed drug development (MIDD) paradigm encourages integration of population PK models into adaptive decisions [18, 19]. Two recent regulatory developments sharpen this expectation further. The ICH M15 guideline on general principles for model-informed drug development, which entered into effect on 23 July 2026, establishes for the first time a harmonised international framework for the prospective planning, model evaluation, and standardised reporting of MIDD evidence submitted to regulatory authorities [20]; by requiring that model assumptions and their impact on decision-making be transparently documented, it raises the evidentiary bar for how pharmacometric information such as PK exposure should be incorporated into adaptive dose-finding decisions. In parallel, the FDA's draft guidance on the use of Bayesian methodology in clinical trials of drug and biological products explicitly identifies dose-finding studies as a setting in which the operating characteristics, prior structure, and decision rules of a Bayesian design should be pre-specified and evaluated before trial conduct [21]. Among model-assisted designs, gBOIN [22] accommodates continuous PK-type endpoints, and Takeda et al. [23] introduce PK-weighted randomisation in a single-agent BOIN variant. Most directly relevant is PKBOIN-12 [24], which embeds a continuous PK exposure metric into the BOIN12 framework, using a PK-sufficiency threshold $\zeta_1 = 0.8\, r_P$ (where $r_P$ is the target exposure) to expand the escalation candidate set when exposure is adequate and to eliminate sub-therapeutic doses at trial completion.

Drug-combination strategies are a mainstay of modern cancer treatment, and identifying the optimal biological dose combination (OBDC) in early-phase combination trials is both statistically and operationally more demanding than single-agent dose-finding, owing to limited per-cell sample sizes and a substantially larger dose-combination search space [25, 26]. Critically, combining the individually optimal doses of two component agents need not yield a globally optimal combination, and may instead produce excessive toxicity or a pharmacokinetic profile that differs materially from what either agent's monotherapy data would predict. Several phase I/II designs have been developed for early-phase drug-combination trials [27, 10, 28, 29], but none extends PK-informed dose-finding methods to the combination setting, and none of the two-dimensional dose-ordering devices used elsewhere in the combination dose-finding literature – for instance the standardised, magnitude-aware dose scores of Thall et al. [30] or the coordinatewise partial orders used in partial-order continual-reassessment designs [31, 32] – has previously been adapted to govern PK-based decisions.

We propose **PKComb-BOIN12**, a pharmacometrically informed, utility-based Bayesian optimal interval design for identifying the OBDC in two-agent early-phase oncology trials. PKComb-BOIN12 embeds a continuous PK exposure metric into the Comb-BOIN12 interval framework: combinations with insufficient PK exposure are excluded from the final admissible set, and the selected OBDC is the admissible combination maximising quasi-beta-binomial posterior mean utility, without fitting a separate parametric dose–response model. Two methodological features distinguish PKComb-BOIN12 from a direct combination-space translation of PKBOIN-12's single-agent mechanisms, and both are developed with a companion theoretical justification rather than as heuristic design choices.

First, because the two-dimensional dose-combination space admits only a partial order under the coordinatewise monotonicity assumed for PK exposure, PK-based elimination cannot, without further justification, be organised around a total order such as a scalar dose-intensity index. We instead define PK elimination through a coordinatewise order-ideal (down-set) removal: evidence that a combination $(j, k)$ has insufficient exposure licenses simultaneous elimination of exactly those combinations that $(j, k)$ coordinatewise dominates from below, $\{(j', k') : j' \leq j,\ k' \leq k\}$, mirroring the up-set removal already used by the safety rule and requiring no auxiliary assumption beyond the monotonicity already invoked for the exposure surface itself (Section 3.2, Proposition 2).

Second, a target exposure informed primarily by monotherapy pharmacokinetics need not retain its clinical meaning once the two agents are co-administered: drug–drug interactions, synergistic or antagonistic pharmacodynamic effects, and shared elimination pathways can shift the combination-arm exposure–response relationship away from what single-agent data would predict [33]. Rather than treating the monotherapy-derived target as fixed throughout the trial, PKComb-BOIN12 treats it as a *transportable-but-verifiable* quantity, updated during the trial by a hierarchical calibration mechanism that shrinks the assumed target toward accruing combination-arm PK data, together with a formal bound on the resulting bias in OBDC selection when transportability is violated (Section 3.1.1, Proposition 1).

A time-to-event extension, TITE-PKComb-BOIN12, further accommodates delayed toxicity and efficacy outcomes by exploiting the earlier availability of PK data relative to long-term clinical endpoints, shortening trial duration without materially eroding selection accuracy. Across a simulation study spanning structurally diverse toxicity–efficacy–PK scenarios, PKComb-BOIN12 improves OBDC selection in the large majority of scenarios in which a true OBDC exists, and TITE-PKComb-BOIN12 further reduces trial duration; a dedicated factorial decomposition separates the contribution of PK-guided real-time escalation from that of restricting the final admissible set on PK grounds and from that of the final-selection estimator, so that the source of these gains is transparent rather than confounded with a change in the target estimand. To facilitate practical implementation, we provide an interactive `R` Shiny application, `PKCombBOIN12`, supporting prior design evaluation, operating-characteristic simulation, and real-time dose-decision support during trial conduct.

The remainder of this paper is organised as follows. Section 2 reviews Comb-BOIN12 and PKBOIN-12. Section 3 presents the proposed PKComb-BOIN12 design, its theoretical properties, and its time-to-event extension. Section 4 reports an extensive simulation study, including a scenario informed by a published combination trial and a factorial decomposition of the sources of PKComb-BOIN12's gains. Section 5 illustrates the design through a retrospective case study. Section 6 provides implementation guidance and describes the `R` Shiny application, and Section 7.1 concludes with directions for future research.

# 2 Background

## 2.1 Optimal Biological Dose Combination

Consider an early-phase trial of two agents, with $J$ pre-specified dose levels of agent A and $K$ pre-specified dose levels of agent B, indexed by $j = 1, \dots, J$ and $k = 1, \dots, K$. A dose combination is denoted $(j, k)$, with actual dose values $a_j$ and $b_k$. The trial collects a binary toxicity outcome $Y_T$ and a binary efficacy outcome $Y_E$, where $Y_T = 1$ indicates a dose-limiting toxicity (DLT) and $Y_E = 1$ represents an efficacy response. Let $p_T$ and $q_E$ denote the maximum acceptable toxicity probability and the minimum acceptable efficacy probability, and let $p_{j,k}$ and $q_{j,k}$ denote the true toxicity and efficacy probabilities at $(j, k)$. As is standard for combination trials, $p_{j,k}$ is assumed non-decreasing in both $j$ and $k$; this assumption is not imposed on $q_{j,k}$, which may be unimodal or non-monotone, particularly for targeted and immuno-oncology agents.

Outcomes are categorised as $O_1 = (Y_T = 0, Y_E = 1)$, $O_2 = (Y_T = 0, Y_E = 0)$, $O_3 = (Y_T = 1, Y_E = 1)$, $O_4 = (Y_T = 1, Y_E = 0)$, with patient counts $n_{j,k,i}$, $i = 1, \dots, 4$, and total sample size $n_{j,k} = \sum_i n_{j,k,i}$ at $(j, k)$. Let $u_i \in [0, 100]$ be a pre-specified utility score for $O_i$, anchored at $u_1 = 100$ and $u_4 = 0$, with $\boldsymbol{u} = (u_1, u_2, u_3, u_4)$. Given the outcome probability vector $\boldsymbol{\pi}_{j,k}$, the expected utility is $EU_{j,k} = \sum_i u_i \pi_{j,k,i}$; when $u_2 + u_3 = 100$, $EU_{j,k} = u_2(1 - p_{j,k}) + u_3\, q_{j,k}$. Following BOIN12 and Comb-BOIN12, we set $(u_2, u_3) = (40, 60)$ throughout, though the framework does not require $u_2 + u_3 = 100$. Without a PK constraint, the OBDC is the admissible combination ($p_{j,k} \le p_T$, $q_{j,k} \ge q_E$) with the highest $EU_{j,k}$; if no combination is admissible, the OBDC does not exist and the trial should terminate early. Section 3.1 extends this definition to incorporate PK sufficiency once the relevant notation has been introduced.

## 2.2 Comb-BOIN12

Comb-BOIN12 [10] extends BOIN12 [9] to two-agent combinations while retaining the quasi-beta-binomial utility model and the RDS-based decision rule. Writing $u^*_{j,k} = EU_{j,k}/100$ for the standardised desirability, the quasi-binomial event count is $x_{j,k} = 100^{-1} \sum_i u_i n_{j,k,i}$, interpretable as a quasi-binomial outcome from $n_{j,k}$ patients, $x_{j,k} \sim \text{quasi-binomial}(n_{j,k}, u^*_{j,k})$. Under a non-informative Beta(1, 1) prior, $u^*_{j,k} \mid n_{j,k}, x_{j,k} \sim \text{Beta}(1 + x_{j,k},\ 1 + n_{j,k} - x_{j,k})$, and the RDS ranks combinations by the posterior tail probability $P(u^*_{j,k} > u_b \mid n_{j,k}, x_{j,k})$ against the benchmark $u_b = \{\underline{u} + (100 - \underline{u})/2\}/100$, with $\underline{u} = u_1(1 - p_T)q_E + u_2(1 - p_T)(1 - q_E) + u_3 p_T q_E$.

Let $\lambda_1, \lambda_2$ denote the BOIN escalation and de-escalation boundaries derived from $p_T$ under the default specification of Liu, Yuan, and Lin [8], and let $N^*$ denote a per-cell sample-size cutoff. Given the current combination $(j, k)$ with observed toxicity rate $\hat{p}_{j,k}$, Comb-BOIN12 selects the next cohort's dose by comparing $\hat{p}_{j,k}$ with $\lambda_1, \lambda_2$: it de-escalates when $\hat{p}_{j,k} \ge \lambda_2$, restricts the candidate set to $\{(j - 1, k), (j, k), (j, k - 1)\}$ when $\hat{p}_{j,k}$ is intermediate and $n_{j,k} \ge N^*$, and otherwise permits movement to any of the four adjacent neighbours, in every case selecting the candidate with the largest RDS. Safety elimination removes $(j, k)$ together with every $(j', k')$ with $j' \ge j, k' \ge k$ once $P(p_{j,k} > p_T \mid \cdot) > C_T$; efficacy (futility) elimination removes $(j, k)$ alone once $P(q_{j,k} < q_E \mid \cdot) > C_E$. At trial completion, Comb-BOIN12 determines the maximum tolerated dose combination (MTDC) via bivariate isotonic regression (pool-adjacent-violators, PAVA [34]) on the observed toxicity matrix, fits a separate parametric efficacy model, and selects the admissible combination with the highest estimated utility; we refer readers to Lu, Zhang, Yuan, and Lin [10] for the complete selection procedure. Comb-BOIN12 substantially advances toxicity-only combination designs but does not incorporate PK information at any stage.

## 2.3 PKBOIN-12

PKBOIN-12 [24] extends BOIN12 to incorporate a continuous PK outcome alongside toxicity and efficacy for single-agent dose-finding. Let $d \in \{1, \dots, D\}$ index single-agent dose levels, with $r_d$ the true mean of a continuous PK metric assumed to increase strictly with dose. PKBOIN-12 places a truncated-normal prior on each $r_d$ and derives the PK-sufficiency cutoff $\zeta_1 = (r_P + r_I)/2 = 0.8\, r_P$, where $r_P$ is the target PK value and $r_I = 0.6\, r_P$ a pre-specified inefficacious value [22, 23]; this cutoff minimises the posterior probability of an incorrect binary PK classification under equal prior probabilities on $H_{P0} : r_d = r_P$ versus $H_{P1} : r_d = r_I$. When the observed PK mean at the current dose exceeds $\zeta_1$, PKBOIN-12 expands the admissible dosing set to include lower dose levels that retain adequate exposure but carry lower toxicity risk; a complementary elimination criterion removes dose levels with high posterior probability of exposure below $r_P$, and the final admissible set is restricted to dose levels at or above the isotonically estimated minimum efficacious PK dose. PKBOIN-12 and Comb-BOIN12 thus address complementary limitations of BOIN12 – single-agent PK incorporation without a combination extension, and combination dose-finding without PK incorporation, respectively – motivating the combination design developed in Section 3.

# 3 Proposed PKComb-BOIN12 Design

## 3.1 Incorporation of PK Outcomes and the PK-Sufficiency Estimand

Many continuous PK outcomes are routinely collected in early-phase combination trials, including the area under the concentration–time curve ($AUC$), maximum concentration ($C_{\max}$), and clearance ($CL$). Following Sun and Tu [24], we focus on a single continuous PK metric presumed to correlate positively with both toxicity and efficacy at the individual level: at a fixed dose combination, patients with higher exposure are more likely to experience both a DLT and an efficacy response. Let $r_{j,k}$ denote the true mean PK outcome at $(j,k)$.

**Assumption 1 (Coordinatewise PK monotonicity).** $r_{j,k}$ is non-decreasing in both $j$ and $k$: for all $j' \geq j$ and $k' \geq k$, $r_{j',k} \geq r_{j,k}$ and $r_{j,k'} \geq r_{j,k}$.

Assumption 1 is most credible when the two agents are metabolised through distinct pathways and do not meaningfully alter one another's absorption or clearance. This is not guaranteed in oncology combinations: ribociclib is a CYP3A4 inhibitor and everolimus a sensitive CYP3A4 substrate, and their co-administration was shown to increase everolimus exposure 1.5- to 3-fold relative to monotherapy in a Phase Ib dose-escalation trial, materially distorting the expected dose–exposure ordering across the grid [35]. The ICH M12 guideline on drug-interaction studies formalises the pre-trial evaluation that should precede reliance on Assumption 1: in vitro assessment of each agent's inhibition, induction, and substrate liabilities, escalating where warranted to a dedicated clinical drug–drug interaction (DDI) assessment or a physiologically based pharmacokinetic (PBPK) prediction [33]. We recommend that investigators review monotherapy PK profiles and known interaction liabilities before applying PKComb-BOIN12, and, where early combination PK data are available, verify that observed exposures are broadly consistent with Assumption 1 before proceeding to dose escalation; Section 3.1.1 develops a formal mechanism for the closely related, but distinct, question of whether the *target level* of exposure transports from monotherapy to the combination.

Under the Bayesian framework, each dose-specific mean $r_{j,k}$ is assigned a truncated-normal prior left-truncated at zero, truncated-$N(0,\sigma_0^2;0,\infty)$ with $\sigma_0$ large (default $\sigma_0 = 10{,}000$). With $I_{j,k}$ the set of patients assigned to $(j,k)$ and individual-level PK values $r_{j,k,i} \overset{\text{i.i.d.}}{\sim} N(r_{j,k},\sigma_{j,k}^2)$ for $i \in I_{j,k}$, the posterior is

$$r_{j,k} \mid \hat{r}_{j,k},\sigma_0^2 \sim \text{truncated-}N\Big(\text{mean} = \frac{n_{j,k}\hat{r}_{j,k}}{\sigma_{j,k}^2(1/\sigma_0^2+n_{j,k}/\sigma_{j,k}^2)}, \quad \text{var} = \frac{1}{1/\sigma_0^2+n_{j,k}/\sigma_{j,k}^2};\ 0,\infty\Big), \tag{1}$$

where $\hat{r}_{j,k} = n_{j,k}^{-1}\sum_{i\in I_{j,k}} r_{j,k,i}$. For $\sigma_0$ large, the maximum a posteriori estimate is approximately $\hat{r}_{j,k}$, used directly for real-time decisions. The within-combination variance $\sigma_{j,k}^2$ is unknown and unstable to estimate separately in cells of three to nine patients; we adopt the standard population-PK working assumption of an approximately constant coefficient of variation, $\sigma_{j,k}^2 \approx (\text{CV}\times r_{j,k})^2$, approximated at each interim analysis by substituting $\hat{r}_{j,k}$ for $r_{j,k}$, with a pooled-variance or historically fixed alternative available when supported by prior single-agent data (Section 4.6.3 evaluates the operating-characteristic impact of this choice being misspecified).

*Remark 1. The untruncated-normal approximation used throughout the design.* The exact posterior (1) both shrinks $\hat{r}_{j,k}$ toward the prior mean 0 and truncates its support to $(0,\infty)$; every real-time and final PK-based decision this design makes – Rule 1 (Section 3.2), Algorithm 1, the final admissibility filter of Section 3.4, and Propositions 1 and 4 below – instead uses the untruncated form $r_{j,k} \mid \hat{r}_{j,k}, n_{j,k} \approx N(\hat{r}_{j,k},\sigma_{j,k}^2/n_{j,k})$ obtained from (1) as $\sigma_0 \to \infty$, and we now quantify why this substitution is accurate at every sample size this design actually acts on. Writing $m_{j,k} = \hat{r}_{j,k}\{1+\sigma_{j,k}^2/(n_{j,k}\sigma_0^2)\}^{-1}$ for the exact posterior mean in (1), $m_{j,k}$ differs from $\hat{r}_{j,k}$ by a relative amount at most $\sigma_{j,k}^2/(n_{j,k}\sigma_0^2)$; under the constant-CV working assumption above, this is $(\mathrm{CV}\,\hat{r}_{j,k}/\sigma_0)^2/n_{j,k}$, which for the default $\sigma_0 = 10{,}000$, CV $= 0.25$, and the target exposure $r_P = 6{,}000$ used throughout Section 4 is at most 0.75% at the smallest sample size ($n_{j,k} = 3$) at which any PK-based decision is made in this design, and under 0.4% once $n_{j,k} \geq 6$; even at the largest exposure value used anywhere in the scenario library ($r \approx 9{,}000$) it remains under 1.7% at $n_{j,k} = 3$. Separately, the probability mass the untruncated approximation places below zero – the truncation this substitution ignores – is $\Phi(-m_{j,k}/\sqrt{v_{j,k}}) \approx \Phi(-\sqrt{n_{j,k}}/\mathrm{CV})$; for CV $= 0.25$ this is below $2.2 \times 10^{-12}$ already at $n_{j,k} = 3$ and below $6 \times 10^{-23}$ at $n_{j,k} = 6$. Both corrections are therefore negligible at every decision point this design makes, which is why the untruncated approximation, rather than (1) itself, is the object analysed in Section 3.1.1 and Propositions 1 and 4.

#### 3.1.1 *Transportability of the PK Target: A Hierarchical Calibration Mechanism*

Let $r_P^{(0)}$ denote a pre-trial target exposure derived from monotherapy PK data, a population PK model, or a prior exposure–response analysis, and let $r_{j,k}^{\mathrm{mono}}$ denote the corresponding pre-trial, monotherapy-based prediction of the mean exposure profile across the dose-combination grid (e.g. obtained by scaling single-agent exposure by the combination dose of each agent under an assumed additive or independent-elimination model). We do not assume that the combination-arm exposure surface coincides with this prediction; instead we introduce a combination-specific *transportability factor* $\tau > 0$ such that the data-generating exposure surface satisfies $r_{j,k} = \tau\, r_{j,k}^{\mathrm{mono}}$, with $\tau = 1$ corresponding to full transportability of the monotherapy-based prediction and $\tau \neq 1$ representing a systematic DDI-driven or pharmacodynamic shift. The PK-sufficiency threshold used by the design is then

$$\zeta_1(\tau) = 0.8\, r_P^{(0)}\, \tau, \tag{2}$$

generalising the PKBOIN-12 threshold $\zeta_1 = 0.8\, r_P$ [24] (recovered at $\tau = 1$).

A naive design fixes $\tau = 1$ throughout the trial, implicitly assuming full transportability. PKComb-BOIN12 instead offers an *adaptive calibration* option, detailed in Algorithm 1, in which the assumed $\tau$ is updated using the ratio of each patient's observed exposure to the monotherapy-based prediction at their assigned dose, shrunk toward the prior belief $\tau = 1$ with prior effective sample size $n_0$ (a pre-specified SAP parameter reflecting confidence in the monotherapy bridge; $n_0 \to \infty$ recovers the naive fixed-$\tau$ design and $n_0 \to 0$ recovers a fully data-driven estimate).

**Algorithm 1 (Adaptive PK-target calibration).** *At each interim decision point, with $n_0$ a pre-specified prior effective sample size:*

1. *For every patient i enrolled so far with observed PK value $r_i$ at assigned combination $(j_i, k_i)$, compute the individual transportability ratio $\hat{\tau}_i = r_i/r_{j_i,k_i}^{\mathrm{mono}}$.*
2. *Form the precision-weighted estimate $\hat{\tau} = \{n_0 \cdot 1 + n_{\mathrm{PK}} \cdot \bar{\hat{\tau}}\}/(n_0 + n_{\mathrm{PK}})$, where $n_{\mathrm{PK}}$ is the number of patients with an observed PK value and $\bar{\hat{\tau}}$ their mean ratio.*
3. *Set the current decision threshold $\zeta_1(\hat{\tau}) = 0.8\, r_P^{(0)}\, \hat{\tau}$ and use it in place of $\zeta_1$ throughout Sections 3.2–3.4 for that decision point.*

**Lemma 1 (Finite-sample bias and mean-squared error of the calibration estimator).** *Fix a single interim decision point of Algorithm 1 and let $n_{\mathrm{PK}}$ denote the number of patients with an observed PK value at that point; $n_{\mathrm{PK}}$ is realised and known once the decision point is reached, and all expectations, variances, and the mean-squared error below are implicitly conditional on this realised value (at a later decision point $n_{\mathrm{PK}}$ will typically have increased and $\hat{\tau}$ is recomputed afresh from the enlarged sample, but within any one evaluation of the algorithm $n_{\mathrm{PK}}$ is fixed, not random). Let $\{\mathcal{F}_i\}_{i=0}^{n_{\mathrm{PK}}}$ be the filtration with $\mathcal{F}_{i-1}$ the σ-field generated by all trial data (allocation and outcome) available immediately before patient i's dose $(j_i, k_i)$ is assigned, so that $(j_i, k_i)$ is $\mathcal{F}_{i-1}$-measurable, as holds under any non-anticipating adaptive allocation rule, including PKComb-BOIN12's; write $\mathcal{F}_i = \sigma(\mathcal{F}_{i-1}, r_i)$ for the enlarged σ-field that additionally reveals patient* *i's own PK outcome $r_i$. Suppose the data-generating exposure surface satisfies $r_{j,k} = \tau_{\text{true}}\, r^{\text{mono}}_{j,k}$ for every $(j,k)$ with $r^{\text{mono}}_{j,k} > 0$ (Section 3.1.1), and that patient i's PK outcome satisfies $r_i \mid \mathcal{F}_{i-1} \sim N(r_{j_i,k_i}, \sigma^2_{j_i,k_i})$ with $0 < \sigma^2_{j_i,k_i} < \infty$ as in Section 3.1. Let $\hat{\tau}$ denote the step-2 estimator of Algorithm 1 based on these $n_{\text{PK}}$ patients, write $\hat{\tau}_i = r_i / r^{\text{mono}}_{j_i,k_i}$ for the individual ratio, and write $\sigma^2_{\tau,i} = \sigma^2_{j_i,k_i}/(r^{\text{mono}}_{j_i,k_i})^2 < \infty$ for its conditional variance. Then:*

(a) $\{\hat{\tau}_i - \tau_{\text{true}}\}_{i=1}^{n_{\text{PK}}}$ *is a martingale difference sequence with respect to $\{\mathcal{F}_i\}$ – that is, $\hat{\tau}_i - \tau_{\text{true}}$ is integrable and $\mathcal{F}_i$-measurable, and $\mathrm{E}(\hat{\tau}_i - \tau_{\text{true}} \mid \mathcal{F}_{i-1}) = 0$ for every i, regardless of which combination is assigned. Consequently*

$$\mathrm{E}(\hat{\tau}) - \tau_{\text{true}} = \frac{n_0}{n_0 + n_{\text{PK}}}(1 - \tau_{\text{true}}); \tag{3}$$

(b) $\mathrm{Var}(\hat{\tau}) = (n_0 + n_{\text{PK}})^{-2} \sum_{i=1}^{n_{\text{PK}}} \mathrm{E}(\sigma^2_{\tau,i})$;

(c) *under the constant-coefficient-of-variation working assumption of Section 3.1 ($\sigma_{j,k} \approx \mathrm{CV} \times r_{j,k}$), $\sigma^2_{\tau,i} \approx (\mathrm{CV} \times \tau_{\text{true}})^2 =: \sigma^2_\tau$ for every i and every assigned combination, so*

$$\mathrm{MSE}(\hat{\tau}) = \left\{\frac{n_0(1 - \tau_{\text{true}})}{n_0 + n_{\text{PK}}}\right\}^2 + \frac{n_{\text{PK}}\, \sigma^2_\tau}{(n_0 + n_{\text{PK}})^2}. \tag{4}$$

*Proof.* (a) *Adaptedness and integrability.* Since $(j_i, k_i)$ is $\mathcal{F}_{i-1}$-measurable and hence $\mathcal{F}_i$-measurable, and $r_i$ is $\mathcal{F}_i$-measurable by construction of $\mathcal{F}_i$, the ratio $\hat{\tau}_i = r_i / r^{\text{mono}}_{j_i,k_i}$ is $\mathcal{F}_i$-measurable as a quotient of $\mathcal{F}_i$-measurable functions with a denominator bounded away from zero ($r^{\text{mono}}_{j_i,k_i} \geq \min_{(j,k)} r^{\text{mono}}_{j,k} > 0$ on the finite grid, by hypothesis). Since $r_i \mid \mathcal{F}_{i-1} \sim N(r_{j_i,k_i}, \sigma^2_{j_i,k_i})$ has finite conditional mean, $\mathrm{E}(|r_i| \mid \mathcal{F}_{i-1}) < \infty$ a.s., so $\mathrm{E}|\hat{\tau}_i| \leq \mathrm{E}(|r_i|)/\min_{(j,k)} r^{\text{mono}}_{j,k} < \infty$ and $\hat{\tau}_i$, hence $\hat{\tau}_i - \tau_{\text{true}}$, is integrable.

*Conditional mean zero.* The denominator $r^{\text{mono}}_{j_i,k_i}$ is a deterministic function of the $\mathcal{F}_{i-1}$-measurable pair $(j_i, k_i)$ and is therefore itself $\mathcal{F}_{i-1}$-measurable, so by the "pull-out" property of conditional expectation,

$$\mathrm{E}(\hat{\tau}_i \mid \mathcal{F}_{i-1}) = \frac{\mathrm{E}(r_i \mid \mathcal{F}_{i-1})}{r^{\text{mono}}_{j_i,k_i}} = \frac{r_{j_i,k_i}}{r^{\text{mono}}_{j_i,k_i}} = \tau_{\text{true}},$$

where the second equality is the conditional-normal model's mean and the third is the transportability model evaluated at whichever combination $(j_i, k_i)$ happens to be $\mathcal{F}_{i-1}$-measurably assigned. Because the same $\tau_{\text{true}}$ multiplies $r^{\text{mono}}_{j,k}$ identically at *every* $(j,k)$ on the grid, this conditional mean equals the constant $\tau_{\text{true}}$ irrespective of the (possibly complex, outcome-dependent) allocation history encoded in $\mathcal{F}_{i-1}$: the identity holds pathwise, not merely on average over allocation histories. This proves $\mathrm{E}(\hat{\tau}_i - \tau_{\text{true}} \mid \mathcal{F}_{i-1}) = 0$ for every $i$, establishing the martingale difference property; the tower property then gives the unconditional mean $\mathrm{E}(\hat{\tau}_i) = \mathrm{E}\{\mathrm{E}(\hat{\tau}_i \mid \mathcal{F}_{i-1})\} = \tau_{\text{true}}$ for every $i = 1, \ldots, n_{\text{PK}}$.

*Bias of $\hat{\tau}$.* With $n_{\text{PK}}$ fixed as stipulated above, $\hat{\tau} = \{n_0 \cdot 1 + n_{\text{PK}}\bar{\hat{\tau}}\}/(n_0 + n_{\text{PK}})$, where $\bar{\hat{\tau}} = n_{\text{PK}}^{-1} \sum_{i=1}^{n_{\text{PK}}} \hat{\tau}_i$, is an affine function of the (finite) sum $\sum_{i=1}^{n_{\text{PK}}} \hat{\tau}_i$; by linearity of expectation and $\mathrm{E}(\hat{\tau}_i) = \tau_{\text{true}}$ for every $i$,

$$\mathrm{E}(\hat{\tau}) = \frac{n_0 + n_{\text{PK}}\, \tau_{\text{true}}}{n_0 + n_{\text{PK}}}.$$

Subtracting $\tau_{\text{true}} = (n_0 + n_{\text{PK}})\tau_{\text{true}}/(n_0 + n_{\text{PK}})$ from both sides and simplifying the numerator, $n_0 + n_{\text{PK}}\tau_{\text{true}} - (n_0 + n_{\text{PK}})\tau_{\text{true}} = n_0(1 - \tau_{\text{true}})$, gives (3).

(b) *Uncorrelated increments.* For $1 \leq l < i \leq n_{\text{PK}}$, $\hat{\tau}_l - \tau_{\text{true}}$ is $\mathcal{F}_l$-measurable and $\mathcal{F}_l \subseteq \mathcal{F}_{i-1}$ (the filtration is increasing), so it may be pulled out of a conditional expectation given $\mathcal{F}_{i-1}$; by the tower property,

$$\begin{aligned}\mathrm{E}\{(\hat{\tau}_i - \tau_{\text{true}})(\hat{\tau}_l - \tau_{\text{true}})\} &= \mathrm{E}\Big[(\hat{\tau}_l - \tau_{\text{true}})\, \mathrm{E}\{\hat{\tau}_i - \tau_{\text{true}} \mid \mathcal{F}_{i-1}\}\Big] \\ &= \mathrm{E}\{(\hat{\tau}_l - \tau_{\text{true}}) \cdot 0\} = 0\end{aligned}$$

by part (a). Hence $\mathrm{Cov}(\hat{\tau}_i, \hat{\tau}_l) = 0$ for $i \neq l$, and the variance of the sum is the sum of the variances:

$$\mathrm{Var}\Big(\sum_{i=1}^{n_{\mathrm{PK}}} \hat{\tau}_i\Big) = \sum_{i=1}^{n_{\mathrm{PK}}} \mathrm{Var}(\hat{\tau}_i).$$

By the law of total variance, $\mathrm{Var}(\hat{\tau}_i) = \mathrm{E}\{\mathrm{Var}(\hat{\tau}_i \mid \mathcal{F}_{i-1})\} + \mathrm{Var}\{\mathrm{E}(\hat{\tau}_i \mid \mathcal{F}_{i-1})\}$. The second term vanishes because part (a) shows $\mathrm{E}(\hat{\tau}_i \mid \mathcal{F}_{i-1}) \equiv \tau_{\mathrm{true}}$ is (a.s.) the same constant regardless of the conditioning history, so it has zero variance; for the first term, $\mathrm{Var}(\hat{\tau}_i \mid \mathcal{F}_{i-1}) = \mathrm{Var}(r_i \mid \mathcal{F}_{i-1})/(r^{\mathrm{mono}}_{j_i,k_i})^2 = \sigma^2_{j_i,k_i}/(r^{\mathrm{mono}}_{j_i,k_i})^2 = \sigma^2_{\tau,i}$, using that $r^{\mathrm{mono}}_{j_i,k_i}$ is $\mathcal{F}_{i-1}$-measurable (so scales out of the conditional variance as its square) and the conditional-normal model's variance. Hence $\mathrm{Var}(\hat{\tau}_i) = \mathrm{E}(\sigma^2_{\tau,i})$, and $\mathrm{Var}(n_{\mathrm{PK}}\bar{\hat{\tau}}) = \sum_{i=1}^{n_{\mathrm{PK}}} \mathrm{E}(\sigma^2_{\tau,i})$. Finally, $\hat{\tau} = n_0/(n_0 + n_{\mathrm{PK}}) \cdot 1 + \{n_{\mathrm{PK}}/(n_0 + n_{\mathrm{PK}})\}\bar{\hat{\tau}}$ is an affine transformation of $\bar{\hat{\tau}}$ in which the additive constant $n_0/(n_0 + n_{\mathrm{PK}})$ contributes no variance, so

$$\begin{aligned}\mathrm{Var}(\hat{\tau}) &= \left(\frac{n_{\mathrm{PK}}}{n_0 + n_{\mathrm{PK}}}\right)^2 \mathrm{Var}(\bar{\hat{\tau}}) = \frac{1}{(n_0 + n_{\mathrm{PK}})^2}\mathrm{Var}(n_{\mathrm{PK}}\bar{\hat{\tau}}) \\ &= \frac{1}{(n_0 + n_{\mathrm{PK}})^2}\sum_{i=1}^{n_{\mathrm{PK}}} \mathrm{E}(\sigma^2_{\tau,i}),\end{aligned}$$

as claimed, using $\mathrm{Var}(\bar{\hat{\tau}}) = \mathrm{Var}(n_{\mathrm{PK}}\bar{\hat{\tau}})/n^2_{\mathrm{PK}}$.

(c) Substituting the constant $\sigma^2_{\tau,i} \approx \sigma^2_\tau$ (not depending on $i$) into (b) gives $\mathrm{Var}(\hat{\tau}) = n_{\mathrm{PK}}\sigma^2_\tau/(n_0 + n_{\mathrm{PK}})^2$. Since mean-squared error decomposes exactly as squared bias plus variance for any estimator, $\mathrm{MSE}(\hat{\tau}) = \mathrm{E}\{(\hat{\tau} - \tau_{\mathrm{true}})^2\} = \{\mathrm{E}(\hat{\tau}) - \tau_{\mathrm{true}}\}^2 + \mathrm{Var}(\hat{\tau})$; substituting (3) for the first term and the preceding display for the second gives (4). □

Lemma 1 makes explicit a feature of Algorithm 1 that is not obvious from its description alone: because the transportability factor is assumed common across the whole combination grid, $\hat{\tau}_i$ is conditionally unbiased for $\tau_{\mathrm{true}}$ whichever combination patient $i$ happens to be assigned to, so the adaptivity of the underlying dose-finding rule – which governs assignment through toxicity, efficacy, and PK data jointly – does not itself introduce bias into the calibration estimator; the only bias is the deliberate, quantified shrinkage toward the prior value $\tau = 1$ in (3), which is $O(n_0/n_{\mathrm{PK}})$ and vanishes as $n_{\mathrm{PK}} \to \infty$ for any fixed $n_0$, while $\mathrm{Var}(\hat{\tau}) = O(1/n_{\mathrm{PK}})$ vanishes at the usual parametric rate. Equation (4) exhibits the bias–variance trade-off governed by $n_0$: a larger $n_0$ shrinks $\hat{\tau}$ more aggressively toward 1, reducing variance early in the trial (when $n_{\mathrm{PK}}$ is small and $\bar{\hat{\tau}}$ is unstable) at the cost of persistent bias if $\tau_{\mathrm{true}} \neq 1$, whereas $n_0 \to 0$ removes the bias term entirely but inflates variance when few combination-arm PK observations have accrued; Section 4.6.4 fixes $n_0 = 12$ (four cohorts' worth of PK information under this design's default cohort size of three patients), balancing these two considerations in the simulation study.

**Proposition 1 (Bound on transportability bias).** *Fix a combination $(j,k)$ at a decision point with $n_{j,k} = n \geq 1$ observed PK values, sample mean $\hat{r}_{j,k}$, and working standard deviation $\sigma_{j,k} > 0$ as in Section 3.1, and let $\zeta_1^{\mathrm{used}} = 0.8\, r_P^{(0)}\tau_{\mathrm{used}}$ and $\zeta_1^{\mathrm{true}} = 0.8\, r_P^{(0)}\tau_{\mathrm{true}}$ denote the PK-sufficiency thresholds (2) that would be applied under an assumed and the true transportability factor respectively, with $\sigma_{j,k}$ and $\hat{r}_{j,k}$ held fixed across the comparison (only the threshold, not the observed data, differs between the two). Under the untruncated-normal approximation to the posterior (1) of $r_{j,k}$ quantified in Remark 1 and used throughout this design and in the proof of Proposition 4, i.e. $r_{j,k} \mid \hat{r}_{j,k}, n$ approximately $N(\hat{r}_{j,k}, \sigma^2_{j,k}/n)$, define for $\zeta \in \mathbb{R}$ the posterior PK-insufficiency probability*

$$h(\zeta) = \mathrm{P}\{r_{j,k} < \zeta \mid \hat{r}_{j,k}, n\} = \Phi\left(\frac{(\zeta - \hat{r}_{j,k})\sqrt{n}}{\sigma_{j,k}}\right), \tag{5}$$

*the quantity Rule 1 compares to $C_P$. Then*

$$\left| h(\zeta_1^{\text{used}}) - h(\zeta_1^{\text{true}}) \right| \le \frac{\sqrt{n}}{\sigma_{j,k}\sqrt{2\pi}} \times 0.8\, r_P^{(0)}\, |\tau_{\text{used}} - \tau_{\text{true}}|. \tag{6}$$

*Proof.* Write $a = \sqrt{n}/\sigma_{j,k} > 0$, so that $h(\zeta) = \Phi\{a(\zeta - \hat{r}_{j,k})\}$ from (5), and let $\phi = \Phi'$ denote the standard normal density.

*Step 1: a uniform bound on $h'$.* Since $\Phi$ is differentiable on $\mathbb{R}$ with $\Phi' = \phi$, the chain rule gives, for every $\zeta \in \mathbb{R}$,

$$h'(\zeta) = \phi\Big\{a(\zeta - \hat{r}_{j,k})\Big\} \cdot a.$$

Because $a > 0$, bounding $h'(\zeta)$ uniformly in $\zeta$ reduces to bounding $\phi(x) = (2\pi)^{-1/2}e^{-x^2/2}$ uniformly in $x \in \mathbb{R}$. Now $\phi$ is differentiable with $\phi'(x) = -x\phi(x)$, so $\phi'(x) > 0$ for $x < 0$, $\phi'(0) = 0$, and $\phi'(x) < 0$ for $x > 0$: $\phi$ is strictly increasing on $(-\infty, 0)$ and strictly decreasing on $(0, \infty)$, so it attains its unique global maximum at $x = 0$, where $\phi(0) = 1/\sqrt{2\pi}$. Hence $\sup_{x\in\mathbb{R}} \phi(x) = 1/\sqrt{2\pi}$, and therefore

$$\sup_{\zeta\in\mathbb{R}} h'(\zeta) = a \sup_{x\in\mathbb{R}} \phi(x) = \frac{\sqrt{n}}{\sigma_{j,k}\sqrt{2\pi}} =: L. \tag{7}$$

*Step 2: Mean Value Theorem.* Without loss of generality suppose $\zeta_1^{\text{used}} \le \zeta_1^{\text{true}}$ (the argument is symmetric otherwise, with the roles of the two thresholds exchanged and the same bound $L$ applying to $|h(\zeta_1^{\text{true}}) - h(\zeta_1^{\text{used}})|$). The function $h$ is continuous on the closed interval $[\zeta_1^{\text{used}}, \zeta_1^{\text{true}}]$ and differentiable on the open interval $(\zeta_1^{\text{used}}, \zeta_1^{\text{true}})$, being a composition of the everywhere-differentiable functions $\Phi$ and $\zeta \mapsto a(\zeta - \hat{r}_{j,k})$; if $\zeta_1^{\text{used}} = \zeta_1^{\text{true}}$ the bound (6) is immediate, so assume $\zeta_1^{\text{used}} < \zeta_1^{\text{true}}$. By the Mean Value Theorem there exists $\xi \in (\zeta_1^{\text{used}}, \zeta_1^{\text{true}})$ such that

$$h(\zeta_1^{\text{true}}) - h(\zeta_1^{\text{used}}) = h'(\xi)\left(\zeta_1^{\text{true}} - \zeta_1^{\text{used}}\right).$$

Since $\zeta_1^{\text{true}} > \zeta_1^{\text{used}}$, the left-hand side is non-negative (indeed $h$ is non-decreasing, as $h' \ge 0$ everywhere), so

$$\left| h(\zeta_1^{\text{true}}) - h(\zeta_1^{\text{used}}) \right| = h'(\xi)\left|\zeta_1^{\text{true}} - \zeta_1^{\text{used}}\right| \le L\left|\zeta_1^{\text{used}} - \zeta_1^{\text{true}}\right|,$$

using (7) to bound $h'(\xi) \le L$.

*Step 3: substituting the threshold gap.* By the linear-in-$\tau$ definition of the PK-sufficiency threshold (2), $\zeta_1^{\text{used}} - \zeta_1^{\text{true}} = 0.8\, r_P^{(0)}(\tau_{\text{used}} - \tau_{\text{true}})$ exactly, so $|\zeta_1^{\text{used}} - \zeta_1^{\text{true}}| = 0.8\, r_P^{(0)}|\tau_{\text{used}} - \tau_{\text{true}}|$. Combining this with Step 2 and $L = \sqrt{n}/(\sigma_{j,k}\sqrt{2\pi})$ from (7) gives (6). □

*Remark 2.* An identical argument, with the roles of $\hat{r}_{j,k}$ and $r_{j,k}$ exchanged, bounds by the same constant $L$ the corresponding *sampling-distribution* quantity $\tilde{h}(\zeta) = \mathrm{P}(\hat{r}_{j,k} < \zeta \mid r_{j,k}) = \Phi\{(\zeta - r_{j,k})\sqrt{n}/\sigma_{j,k}\}$, in which the true exposure $r_{j,k}$ is held fixed and the estimator $\hat{r}_{j,k}$ is treated as random; this is the quantity that governs the frequentist sampling behaviour of a single threshold comparison and is the object used in the large-$n$ argument of Proposition 4. The two probabilities $h$ and $\tilde{h}$ share the identical Lipschitz constant $L = \sqrt{n}/(\sigma_{j,k}\sqrt{2\pi})$ because both reduce, after the same location-scale change of variable, to a bound on $\sup_x \phi(x)$; (6) therefore holds equally well with $h$ replaced by $\tilde{h}$ throughout.

Two consequences of Proposition 1 are practically important. First, the bound is linear in the transportability error $|\tau_{\text{used}} - \tau_{\text{true}}|$, so a design that shrinks $\tau_{\text{used}}$ toward $\tau_{\text{true}}$ using accruing data (Algorithm 1) has a bias bound that shrinks correspondingly; by Jensen's inequality and Lemma 1(c), $\mathrm{E}|\hat{\tau} - \tau_{\text{true}}| \le \sqrt{\mathrm{MSE}(\hat{\tau})}$, and substituting (4) shows this upper bound is $O(n_0/n_{\text{PK}}) + O(n_{\text{PK}}^{-1/2}) = O(n_{\text{PK}}^{-1/2})$ for

fixed $n_0$, so the classification-probability bound in (6) vanishes as combination-arm PK data accrue, whereas it remains fixed at its initial value $0.8\, r_P^{(0)} |\tau_{\text{used}} - \tau_{\text{true}}| \times \sqrt{n}/(\sigma_{j,k}\sqrt{2\pi})$ for a naive fixed-$\tau_{\text{used}} = 1$ design regardless of sample size. Second, the bound is *increasing* in $\sqrt{n}/\sigma_{j,k}$: as more PK data accrue at a fixed, uncorrected threshold, the sampling distribution of $\hat{r}_{j,k}$ concentrates around the true $r_{j,k}$, so a fixed targeting error translates into an increasingly deterministic (and, if uncorrected, increasingly wrong) classification – indeed, Proposition 4 below shows this classification becomes deterministically wrong in the large-sample limit whenever $\tau_{\text{used}}$ places the working threshold on the wrong side of the true $r_{j,k}$. This formalises why transportability cannot be discharged by a one-off sensitivity analysis around a fixed reference value: an uncorrected transportability error does not average out with more data, it hardens, while Algorithm 1's bias vanishes at rate $n_0/n_{\text{PK}}$ and its variance at the usual parametric rate $n_{\text{PK}}^{-1}$. Section 4.6.4 evaluates Algorithm 1 against the naive fixed-target design by simulation under $\tau_{\text{true}} \in \{0.7, 1.0, 1.3\}$.

**Definition 1 (OBDC under toxicity, efficacy, and PK exposure).** Combining the toxicity–efficacy OBDC of Section 2.1 with the PK-sufficiency threshold $\zeta_1(\tau)$ of (2), PKComb-BOIN12 targets

$$(j^*, k^*) = \arg\max_{(j,k)} EU_{j,k}$$
$$\text{subject to} \quad p_{j,k} \le p_T, \;\; q_{j,k} \ge q_E, \;\; r_{j,k} \ge \zeta_1(\tau).$$

If no combination satisfies all three constraints jointly, the OBDC does not exist and the trial should terminate early. When every combination has $r_{j,k} \ge \zeta_1(\tau)$, Definition 1 reduces exactly to the toxicity–efficacy OBDC of Section 2.1.

We emphasise a scope limitation made explicit by Definition 1 and returned to in Section 7: the design presumes that a single continuous exposure metric, monitored on one pre-specified scale, is what determines sufficiency across the whole combination grid. When a clinically effective combination is expected on mechanistic grounds to act through a pathway not well captured by that scalar summary – for example when the two agents' individually sufficient exposures act synergistically at levels that would individually be classified insufficient – Definition 1 is not the correct estimand, and a design that does not condition dose recommendations on PK exposure (Comb-BOIN12) remains the appropriate default; we return to this scenario class in the sensitivity analysis of Section 4.6.2.

### 3.2 PK-Based Elimination: A Partial-Order-Consistent Rule

The two-dimensional dose grid $\{1, \dots, J\} \times \{1, \dots, K\}$ is only partially ordered by the coordinatewise relation $(j', k') \preceq (j, k) \iff j' \le j$ and $k' \le k$. Under Assumption 1, evidence that $r_{j,k} < \zeta_1$ licenses the inference $r_{j',k'} \le r_{j,k} < \zeta_1$ for every $(j', k') \preceq (j, k)$, but licenses *no* inference whatsoever about a combination $(j'', k'')$ that is incomparable to $(j, k)$ under $\preceq$ (neither $(j'', k'') \preceq (j, k)$ nor $(j, k) \preceq (j'', k'')$). PKComb-BOIN12 therefore defines PK elimination as follows.

**Rule 1 (PK elimination).** *If* $\mathrm{P}\{r_{j,k} < \zeta_1(\tau) \mid \hat{r}_{j,k}, n_{j,k}\} > C_P$ *and* $n_{j,k} \ge N^*$, *eliminate the down-set of* $(j, k)$,

$$D(j, k) \;=\; \{(j', k') : j' \le j,\; k' \le k\},$$

*from the dose matrix.*

This mirrors the safety rule's elimination of the up-set $U(j, k) = \{(j', k') : j' \ge j, k' \ge k\}$ once $\mathrm{P}(p_{j,k} > p_T \mid \cdot) > C_T$, which is licensed by the analogous monotonicity assumed for toxicity. Unlike a simpler total-order rule – which eliminates "the currently uneliminated combination with the smallest $j + k$, ties broken by the smaller $j$", regardless of where on the grid the triggering evidence arose (Example 1) – down-set removal requires no auxiliary total order and eliminates only combinations for which Assumption 1 itself supplies the inference.

**Proposition 2 (Soundness of down-set PK elimination).** *Let $E \subseteq \{1,\dots,J\} \times \{1,\dots,K\}$ denote the set of combinations eliminated by the PK-elimination rule at any point during the trial, and let $T \subseteq E$ denote the set of combinations at which the rule was directly triggered (i.e. at which the posterior criterion $\mathrm{P}\{r_{j,k} < \zeta_1 \mid \cdot\} > C_P$ held with $n_{j,k} \geq N^*$). Under Assumption 1:*

(a) *$E = \bigcup_{(j,k)\in T} D(j,k)$ is a down-set of the coordinatewise partial order;*

(b) *every $(j',k') \in E$ is coordinatewise dominated by some $(j,k) \in T$ at which the rule actually triggered, i.e. $(j',k') \preceq (j,k)$ for some $(j,k) \in T$;*

(c) *consequently, for every $(j',k') \in E$, $r_{j',k'} \leq r_{j,k}$ for the corresponding dominating $(j,k) \in T$, so elimination of $(j',k')$ is licensed by Assumption 1 applied to directly observed evidence at $(j,k)$.*

*Proof.* Recall that $\preceq$ is a partial order on $\{1,\dots,J\} \times \{1,\dots,K\}$: it is reflexive ($(j,k) \preceq (j,k)$), antisymmetric ($(j',k') \preceq (j,k)$ and $(j,k) \preceq (j',k')$ imply $j' = j, k' = k$), and transitive ($(j'',k'') \preceq (j',k') \preceq (j,k)$ implies $j'' \leq j' \leq j$ and $k'' \leq k' \leq k$, so $(j'',k'') \preceq (j,k)$), each following directly from the corresponding property of $\leq$ on $\{1,\dots,J\}$ and on $\{1,\dots,K\}$ applied coordinatewise. A set $S$ is a *down-set* of $\preceq$ if $(j',k') \preceq (j,k) \in S$ implies $(j',k') \in S$; each $D(j,k)$ is itself a down-set (the principal order ideal generated by $(j,k)$), since $(j'',k'') \preceq (j',k') \in D(j,k)$ gives $(j'',k'') \preceq (j,k)$ by transitivity, so $(j'',k'') \in D(j,k)$.

(a) We show $E = \bigcup_{(j,k)\in T} D(j,k)$ is a down-set. Let $(j'',k'') \preceq (j',k') \in E$; since $E$ is a union of the sets $D(j,k)$ over $(j,k) \in T$, $(j',k') \in D(j,k)$ for some $(j,k) \in T$, and the previous paragraph gives $(j'',k'') \in D(j,k) \subseteq E$. As $(j'',k'') \preceq (j',k') \in E$ was arbitrary, $E$ is a down-set, proving (a); it is finite as $T$ is a subset of the finite grid, so no issue of infinite unions arises.

(b) Let $(j',k') \in E = \bigcup_{(j,k)\in T} D(j,k)$. By definition of a union, $(j',k') \in D(j,k)$ for at least one $(j,k) \in T$, i.e. $j' \leq j$ and $k' \leq k$ for that $(j,k)$, which is exactly $(j',k') \preceq (j,k)$; this $(j,k) \in T$ is a combination at which the rule directly triggered, proving (b).

(c) Fix $(j',k') \in E$ and let $(j,k) \in T$ be a dominating combination from (b), so $(j',k') \preceq (j,k)$, i.e. $j' \leq j$ and $k' \leq k$. Assumption 1 states that $r_{j,k}$ is non-decreasing in each coordinate separately; applying this twice, first in the row coordinate ($r_{j',k} \leq r_{j,k}$, since $j' \leq j$) and then in the column coordinate ($r_{j',k'} \leq r_{j',k}$, since $k' \leq k$), gives $r_{j',k'} \leq r_{j',k} \leq r_{j,k}$. Because $(j,k) \in T$, the rule's posterior criterion held there, i.e. $\mathrm{P}\{r_{j,k} < \zeta_1 \mid \hat{r}_{j,k}, n_{j,k}\} > C_P$ from data observed directly at $(j,k)$; the chain of inequalities $r_{j',k'} \leq r_{j,k}$ transmits this evidence to $(j',k')$ without requiring any observation at $(j',k')$ itself, so elimination of $(j',k')$ is licensed by Assumption 1 applied to evidence actually observed at $(j,k)$, proving (c). □

Proposition 2 establishes that down-set elimination never removes a combination for reasons Assumption 1 does not support (*soundness*); it does not by itself rule out an alternative rule that eliminates a strictly smaller set on the same evidence, which would be sound but overly conservative, wasting some of the inferential leverage the monotonicity assumption legitimately provides. The next result shows this cannot happen: $D(j,k)$ is not merely *a* sound elimination set, it is *the* set of combinations whose PK insufficiency is a logical consequence of $r_{j,k} < \zeta_1$ under Assumption 1 alone, with no combination omitted and none wrongly included.

**Proposition 3 (Exactness of down-set elimination).** *Write $\mathcal{G} = \{1,\dots,J\} \times \{1,\dots,K\}$ for the dose grid. Fix a triggering combination $(j,k)$ and let*

$$L(j,k) = \Big\{(j',k') \in \mathcal{G} : r_{j',k'} < \zeta_1 \text{ for every } r : \mathcal{G} \to \mathbb{R} \text{ satisfying Assumption 1 and } r_{j,k} < \zeta_1\Big\}$$

*denote the set of combinations whose PK insufficiency is entailed by monotonicity alone, given only the single observation $r_{j,k} < \zeta_1$ at $(j,k)$, with no further information about the surface assumed. Then $L(j,k) = D(j,k)$.*

*Proof.* $D(j,k) \subseteq L(j,k)$: if $(j',k') \in D(j,k)$, the argument in the proof of Proposition 2(c) shows $r_{j',k'} \le r_{j,k} < \zeta_1$ for *any* surface satisfying Assumption 1, so $(j',k') \in L(j,k)$.

$L(j,k) \subseteq D(j,k)$: we show the contrapositive, that $(j',k') \notin D(j,k)$ implies $(j',k') \notin L(j,k)$, by exhibiting a monotone surface consistent with $r_{j,k} < \zeta_1$ under which $r_{j',k'} \ge \zeta_1$. Since $(j',k') \notin D(j,k)$ means $\neg(j' \le j$ and $k' \le k)$, we have $j' > j$ or $k' > k$; suppose first $j' > j$ (the case $k' > k$ is symmetric, exchanging the roles of rows and columns below). Fix any $\epsilon \in (0, \zeta_1)$ and define

$$\tilde{r}_{a,b} = \begin{cases} \zeta_1 - \epsilon, & a \le j, \\ \zeta_1 + \epsilon, & a > j, \end{cases} \qquad a \in \{1, \dots, J\},\ b \in \{1, \dots, K\}.$$

This surface is constant in $b$ for each $a$, hence trivially non-decreasing in the column coordinate; in the row coordinate it takes the smaller value $\zeta_1 - \epsilon$ for $a \le j$ and the larger value $\zeta_1 + \epsilon$ for $a > j$, hence is non-decreasing in $a$, so $\tilde{r}$ satisfies Assumption 1. At $(j,k)$: since $j \le j$, $\tilde{r}_{j,k} = \zeta_1 - \epsilon < \zeta_1$, so this surface is consistent with the premise $r_{j,k} < \zeta_1$. At $(j',k')$: since $j' > j$, $\tilde{r}_{j',k'} = \zeta_1 + \epsilon \ge \zeta_1$, so $(j',k') \notin L(j,k)$ under this surface, and hence $(j',k') \notin L(j,k)$ as defined (which requires membership under *every* admissible surface). The symmetric construction for the case $k' > k$ – setting $\tilde{r}_{a,b} = \zeta_1 - \epsilon$ for $b \le k$ and $\zeta_1 + \epsilon$ for $b > k$, which is non-decreasing in $b$ and constant (hence non-decreasing) in $a$ – completes the proof. □

Proposition 3 makes precise the minimality claim already illustrated in Example 1: eliminating any strict subset of $D(j,k)$, as an alternative rule might, would exclude some $(j',k') \in L(j,k) \setminus D(j,k) = \varnothing$ – vacuously, since the two sets coincide – so no combination whose insufficiency is entailed by Assumption 1 is left unaddressed; conversely, eliminating anything beyond $D(j,k)$, as the total-order rule of Example 1 does, necessarily eliminates some $(j',k') \notin L(j,k)$, i.e. a combination for which a monotone surface exists that is consistent with the triggering evidence yet leaves $(j',k')$ PK-sufficient, which is exactly the failure mode Example 1 exhibits concretely. Down-set removal is therefore the unique elimination rule that is simultaneously sound (Proposition 2) and lossless (Proposition 3) with respect to Assumption 1.

Soundness and exactness describe what a single triggering observation licenses; the following result describes what happens as evidence accumulates, showing that the rule's classification of any individual combination becomes correct with certainty in the large-sample limit, provided that combination's true exposure is not exactly on the decision boundary.

**Proposition 4 (Large-sample consistency of PK elimination).** *Fix a combination $(j,k)$ that continues to receive patients, so that $n_{j,k} \to \infty$, with $\hat{r}_{j,k}$, the posterior of $r_{j,k}$, and $\sigma_{j,k}$ as in Section 3.1 (equation (1), used in its untruncated normal approximation as $\sigma_0 \to \infty$, quantified in Remark 1), and fix $C_P \in (0,1)$ and a threshold $\zeta_1(\tau)$ that either is held fixed or converges almost surely to a fixed limit as $n_{\mathrm{PK}} \to \infty$ (the latter holding under Algorithm 1: over the finite dose grid, $\sigma^2_{\tau,i} \le \bar{\sigma}^2_\tau := \max_{(j,k)} \sigma^2_{j,k}/(r^{\mathrm{mono}}_{j,k})^2 < \infty$ for every $i$ by Lemma 1's construction, so $\sum_{i\ge1} \mathrm{Var}(\hat{\tau}_i)/i^2 \le \bar{\sigma}^2_\tau \sum_{i\ge1} i^{-2} < \infty$, and this Kolmogorov-type condition on the martingale difference sequence $\{\hat{\tau}_i - \tau_{\mathrm{true}}\}$ of Lemma 1(a) is sufficient for the martingale strong law of large numbers to give $\bar{\hat{\tau}} - \tau_{\mathrm{true}} \to 0$ almost surely as $n_{\mathrm{PK}} \to \infty$, and hence, since $n_0$ is fixed, $\hat{\tau} \to \tau_{\mathrm{true}}$ almost surely). Then, almost surely on the event $\{r_{j,k} \ne \zeta_1(\tau)\}$:*

- (a) *if $r_{j,k} < \zeta_1(\tau)$, $\mathrm{P}\{r_{j,k} < \zeta_1(\tau) \mid \hat{r}_{j,k}, n_{j,k}\} \to 1$, so Rule 1 triggers at $(j,k)$ for all sufficiently large $n_{j,k}$;*
- (b) *if $r_{j,k} > \zeta_1(\tau)$, $\mathrm{P}\{r_{j,k} < \zeta_1(\tau) \mid \hat{r}_{j,k}, n_{j,k}\} \to 0$, so Rule 1 fails to trigger at $(j,k)$ for all sufficiently large $n_{j,k}$, and, by Proposition 3, $(j,k)$ is eventually eliminated only if some other, genuinely PK-insufficient combination dominating it under $\preceq$ triggers the rule, never on the strength of evidence at $(j,k)$ itself.*

*Proof.* The individual-level PK values $r_{j,k,i}$, $i \in I_{j,k}$, are i.i.d. $N(r_{j,k}, \sigma^2_{j,k})$ (Section 3.1), so the strong law of large numbers gives $\hat{r}_{j,k} = n^{-1}_{j,k} \sum_{i \in I_{j,k}} r_{j,k,i} \to r_{j,k}$ almost surely as $n_{j,k} \to \infty$. From (1), the exact posterior mean $m_{j,k} = n_{j,k}\hat{r}_{j,k}/\{\sigma^2_{j,k}(1/\sigma^2_0 + n_{j,k}/\sigma^2_{j,k})\}$ and variance $v_{j,k} = \{1/\sigma^2_0 + n_{j,k}/\sigma^2_{j,k}\}^{-1}$ of the untruncated approximation satisfy, for fixed $\sigma_0$ and $\sigma_{j,k}$, $m_{j,k} = \hat{r}_{j,k} \cdot \{1 + \sigma^2_{j,k}/(\sigma^2_0 n_{j,k})\}^{-1} \to \hat{r}_{j,k}$ and $\upsilon_{j,k} = (\sigma^2_{j,k}/n_{j,k}) \cdot \{1 + \sigma^2_{j,k}/(\sigma^2_0 n_{j,k})\}^{-1} = O(1/n_{j,k}) \to 0$ as $n_{j,k} \to \infty$; combined with $\hat{r}_{j,k} \to r_{j,k}$ a.s., this gives $m_{j,k} \to r_{j,k}$ a.s. and $\upsilon_{j,k} \to 0$. This also disposes of the truncation Remark 1 quantifies at finite $n_{j,k}$: since $r_{j,k} > 0$ (Section 3.1) and $m_{j,k} \to r_{j,k} > 0$ a.s. while $\sqrt{\upsilon_{j,k}} \to 0$, the untruncated approximation's mass below zero, $\Phi(-m_{j,k}/\sqrt{\upsilon_{j,k}})$, converges to 0 a.s. as well, so the two approximations coincide in the limit used below. Writing $\Phi$ for the standard normal CDF, the (asymptotically exact) approximate posterior gives $\mathrm{P}\{r_{j,k} < \zeta_1(\tau) \mid \hat{r}_{j,k}, n_{j,k}\} = \Phi\{(\zeta_1(\tau) - m_{j,k})/\sqrt{\upsilon_{j,k}}\}$. On $\{r_{j,k} < \zeta_1(\tau)\}$, $\zeta_1(\tau) - m_{j,k} \to \zeta_1(\tau) - r_{j,k} > 0$ a.s. while $\sqrt{\upsilon_{j,k}} \to 0$, so the argument of $\Phi$ diverges to $+\infty$ a.s. and $\Phi\{\cdot\} \to 1$, proving (a); on $\{r_{j,k} > \zeta_1(\tau)\}$ the argument diverges to $-\infty$ a.s. and $\Phi\{\cdot\} \to 0$, proving the first claim of (b). Since $C_P \in (0, 1)$ is fixed, $\Phi\{\cdot\} > C_P$ eventually a.s. in case (a) and $\Phi\{\cdot\} \le C_P$ eventually a.s. in case (b), which is exactly Rule 1's triggering condition (with $n_{j,k} \ge N^*$ satisfied eventually since $n_{j,k} \to \infty$); the final claim of (b) then follows because Proposition 3 shows $D(j, k)$ is exactly the set of combinations whose elimination the evidence at $(j, k)$ licenses, so once the rule stops triggering at $(j, k)$, $(j, k)$ can only be reached from outside, via $D(j', k')$ for some other triggering $(j', k') \ge (j, k)$. □

Proposition 4 does not guarantee that a *finite*, typically small, early-phase trial reaches the large-sample regime at every combination – per-cell sample sizes of three to nine patients are the norm, not the exception, in the simulation scenarios of Section 4 – but it does establish that the rule's finite-sample conservatism relative to an oracle that knows $r_{j,k}$ exactly (documented empirically in Section 4.6.1) is a property of small samples rather than a structural flaw that persists however much data accrue, which establishes the rule's long-run validity formally rather than merely illustrating it by simulation.

Proposition 2 fails for the simpler total-order rule, as the following explicit configuration shows.

**Example 1 (The total-order rule can eliminate a non-dominated combination).** Consider a $2 \times 4$ grid ($J = 2, K = 4$). Suppose the PK-elimination criterion triggers at $(j, k) = (2, 1)$, so that Assumption 1 licenses elimination of $D(2, 1) = \{(1, 1), (2, 1)\}$ only. Suppose combinations $(1, 1)$, $(1, 2)$, $(2, 1)$, and $(1, 3)$ remain uneliminated at the moment of triggering. Under the total-order rule, the combination actually removed is the uneliminated combination with the smallest $j + k$, ties broken by the smaller $j$: here $(1, 1)$ and $(2, 1)$ both have $j + k = 3$, but $(1, 2)$ also has $j + k = 3$ with the same tie value once $(1, 1)$ is excluded by the tie-break rule favouring smaller $j$ among the $j + k = 3$ combinations – so, depending on which combinations at $j + k = 3$ remain, the rule can remove $(1, 2)$ while leaving $(2, 1)$, the actual triggering combination, untouched. Since $(1, 2)$ and $(2, 1)$ are incomparable under $\preceq$ ($1 < 2$ but $2 > 1$), $(1, 2) \notin D(2, 1)$, and Assumption 1 supplies no inference that $r_{1,2} < \zeta_1$ from evidence observed at $(2, 1)$. The total-order rule can therefore eliminate a combination for reasons the design's own monotonicity assumption does not support.

Figure 1 illustrates the two rules on the configuration of Example 1, and Section 4.6.1 compares their simulated operating characteristics directly.

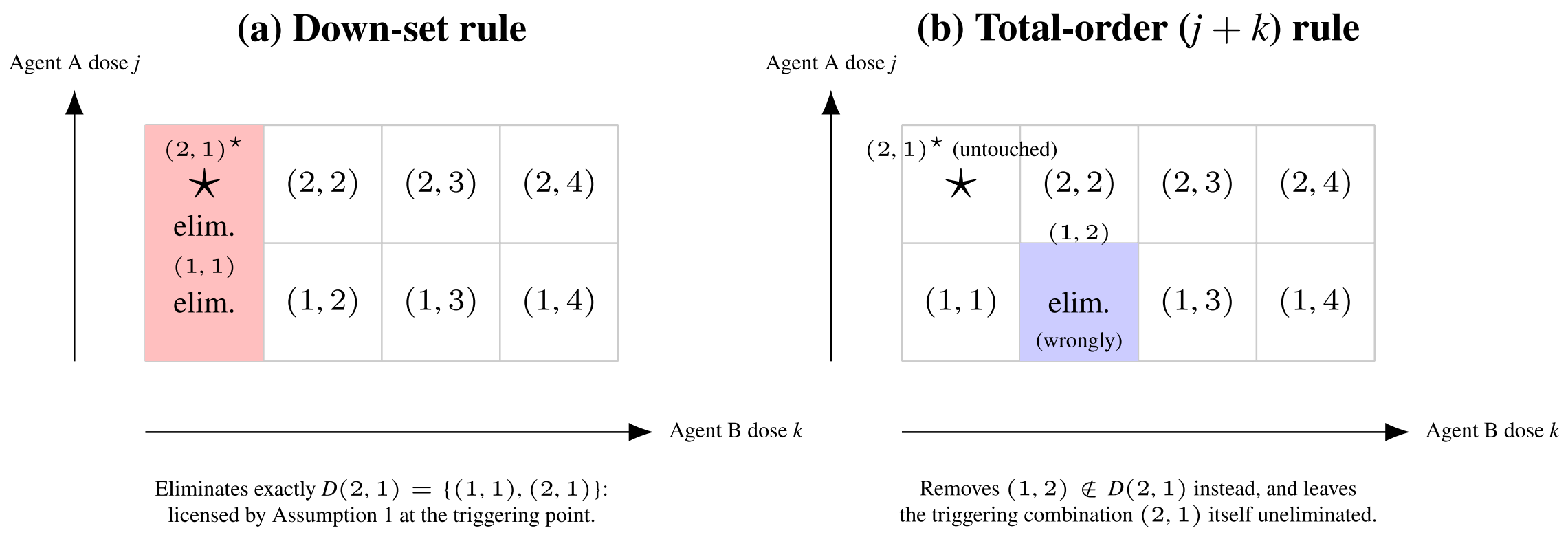

**Figure 1.** Comparison of the proposed down-set PK-elimination rule (a) and the alternative total-order $j+k$ rule (b) on the $2 \times 4$ configuration of Example 1, in which the PK-sufficiency criterion first triggers at the starred combination $(2,1)$. Shaded cells are eliminated. Panel (a) removes exactly the down-set $D(2,1)$, the set of combinations Assumption 1 licenses as PK-insufficient given evidence at $(2,1)$. Panel (b) instead removes $(1,2)$, which is incomparable to $(2,1)$ under $\preceq$ and hence not licensed by the monotonicity assumption, while leaving $(2,1)$ itself in the active set.

### 3.3 Dose-Finding Algorithm

Figure 2 summarises the trial-conduct algorithm described in full below; the corresponding OBDC selection procedure at trial completion is detailed in Section 3.4.

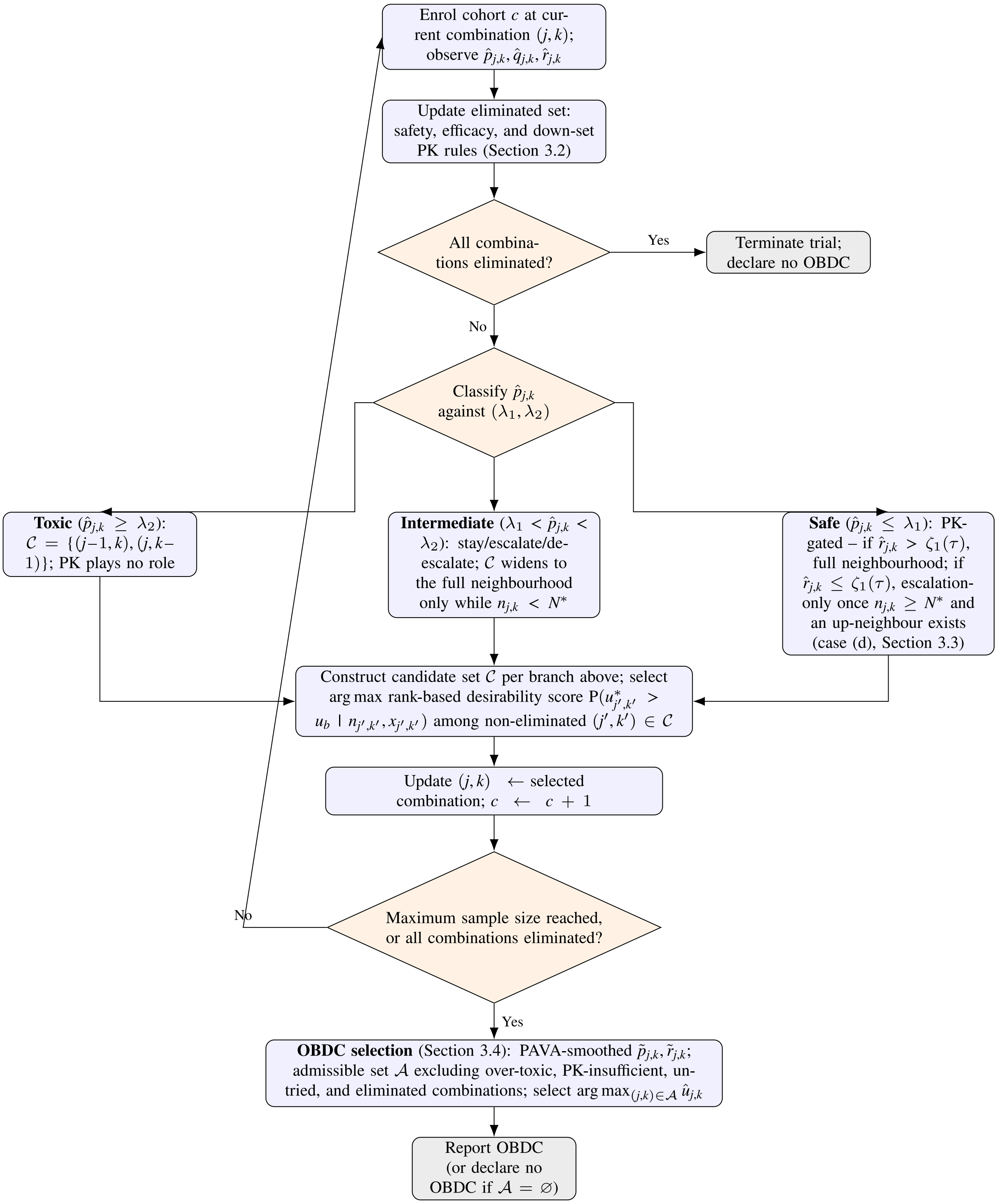

**Figure 2.** Flowchart of the PKComb-BOIN12 trial-conduct algorithm (Section 3.3) and OBDC selection procedure (Section 3.4). PK information enters only through elimination (Section 3.2), the candidate-set construction within the safe branch, and the admissible-set restriction at final selection; escalation and de-escalation decisions in the toxic branch are governed by toxicity and efficacy alone.

The dose-allocation algorithm builds on Comb-BOIN12 by introducing a PK branch that activates only when the current combination is well tolerated, on the principle that PK information is most useful when toxicity does not already constrain the direction of movement. Let $c = 1$ index the initial cohort and $(1, 1)$ the starting combination. At each cohort:

1. Treat cohort $c$ at the current combination $(j, k)$ and derive $\hat{p}_{j,k}$, $\hat{q}_{j,k}$, $\hat{r}_{j,k}$.

2. Apply the safety, efficacy, and PK elimination rules (Section 3.2); if all combinations are eliminated, terminate the trial.
3. Construct the candidate set $\mathcal{C}$ for the next cohort:
   (a) *Toxic* ($\hat{p}_{j,k} \geq \lambda_2$): $\mathcal{C} = \{(j-1,k),(j,k-1)\}$; PK does not enter.
   (b) *Intermediate, saturated* ($\lambda_1 < \hat{p}_{j,k} < \lambda_2$, $n_{j,k} \geq N^*$): $\mathcal{C} = \{(j-1,k),(j,k),(j,k-1)\}$.
   (c) *Intermediate, unsaturated* ($\lambda_1 < \hat{p}_{j,k} < \lambda_2$, $n_{j,k} < N^*$): $\mathcal{C} = \{(j-1,k),(j,k),(j,k-1),(j+1,k),(j,k+1)\}$.
   (d) *Safe* ($\hat{p}_{j,k} \leq \lambda_1$): if $\hat{r}_{j,k} > \zeta_1(\tau)$, $\mathcal{C} = \{(j-1,k),(j,k),(j,k-1),(j+1,k),(j,k+1)\}$. If $\hat{r}_{j,k} \leq \zeta_1(\tau)$, exposure has not reached the active range and de-escalation is excluded: while $n_{j,k}$ is still below one full cohort, $\mathcal{C} = \{(j,k)\} \cup U'(j,k)$ (with $U'(j,k)$ the valid up-neighbours) to avoid forcing escalation on minimal evidence; once $n_{j,k}$ reaches a full cohort and an escalation direction exists, $(j,k)$ is dropped from the candidate set, $\mathcal{C} = U'(j,k)$, so that the RDS comparison genuinely compares escalation options rather than re-confirming a combination already flagged as PK-insufficient.
4. Among combinations in $\mathcal{C}$ not eliminated, select the combination with the largest RDS, $\mathrm{P}(u^*_{j',k'} > u_b \mid n_{j',k'}, x_{j',k'})$; an untried combination is assigned the prior tail probability $1-u_b$.
5. Set $c = c+1$, update $(j,k)$ to the selected combination, and repeat until the maximum sample size is reached or all combinations are eliminated.

The refinement to case (d) relative to a literal reading of the single-agent PKBOIN-12 mechanics – dropping $(j,k)$ from the candidate set once one full cohort of local evidence is available – makes the stated design intent ("escalation is prioritised") an operational property of the algorithm rather than only a qualitative description: retaining $(j,k)$ indefinitely allows the RDS comparison to favour a combination already known to be PK-insufficient whenever it happens to have accumulated a favourable safety/efficacy record, which would silently defeat the PK branch's purpose.

### 3.4 OBDC Selection

At trial completion, we estimate the toxicity, efficacy, and PK surfaces from the accumulated data and select the OBDC as in Definition 1. Bivariate isotonic regression via PAVA [34], weighted by $n_{j,k}$, is applied to $\{\hat{p}_{j,k}\}$ and $\{\hat{r}_{j,k}\}$ to obtain monotone estimates $\{\tilde{p}_{j,k}\}$ and $\{\tilde{r}_{j,k}\}$. Recalling $x_{j,k} = 100^{-1}\sum_i u_i n_{j,k,i}$, the posterior mean standardised desirability $\hat{u}_{j,k} = (1+x_{j,k})/(n_{j,k}+2)$ provides the final-selection utility estimate, reusing the same quasi-beta-binomial posterior that drives the trial-time RDS rather than fitting a separate parametric efficacy model. The OBDC is selected in three steps: (1) determine the MTDC by $(j^*_{\mathrm{MTD}}, k^*_{\mathrm{MTD}}) = \arg\min_{(j,k):n_{j,k}>0} |\tilde{p}_{j,k} - p_T|$; (2) construct the admissible set $\mathcal{A}$ by excluding untreated or eliminated combinations, combinations strictly dominating the MTDC with $\tilde{p}_{j,k} - \tilde{p}_{j^*_{\mathrm{MTD}},k^*_{\mathrm{MTD}}} \geq 0.05$, and combinations with $n_{j,k} \geq 3$ and $0 < \tilde{r}_{j,k} < \zeta_1(\hat{\tau})$ (the requirement $\tilde{r}_{j,k} > 0$ excludes the degenerate case, possible only when $n_{j,k}$ is just above the threshold and every neighbouring untried combination contributes a zero placeholder to the weighted PAVA pool, in which the smoothed estimate itself is uninformative); (3) select $(j^*_{\mathrm{OBDC}}, k^*_{\mathrm{OBDC}}) = \arg\max_{(j,k)\in\mathcal{A}} 100\,\hat{u}_{j,k}$.

**Proposition 5 (Large-sample consistency of the final PK-admissibility filter).** *Suppose every combination $(j,k) \in \mathcal{G}$ receives $n_{j,k} \to \infty$ patients as the trial's planned sample size grows, and Assumption 1 holds. Let $\tilde{r}_{j,k}$ denote the $n_{j,k}$-weighted PAVA isotonic regression estimate of $r_{j,k}$ defined above, and let $\hat{\tau} \to \tau_{\mathrm{true}}$ almost surely as established in the proof of Proposition 4 under Algorithm 1. Then $\tilde{r}_{j,k} \to r_{j,k}$ almost surely for every $(j,k) \in \mathcal{G}$ simultaneously, and consequently, almost surely on $\{r_{j,k} \neq \zeta_1(\tau_{\mathrm{true}})\}$, the final-selection PK-admissibility indicator converges to the correct classification,*

$$\mathbb{1}\{n_{j,k} \geq 3,\ 0 < \tilde{r}_{j,k} < \zeta_1(\hat{\tau})\} \longrightarrow \mathbb{1}\{r_{j,k} < \zeta_1(\tau_{\mathrm{true}})\} \quad \text{a.s.},$$

*i.e. combination $(j, k)$ is correctly retained in, or excluded from, the final admissible set $\mathcal{A}$ for all sufficiently large sample sizes.*

*Proof.* Since $\mathcal{G}$ is finite and $n_{j',k'} \to \infty$ for every $(j', k') \in \mathcal{G}$, the strong law of large numbers gives $\hat{r}_{j',k'} \to r_{j',k'}$ almost surely for every $(j', k') \in \mathcal{G}$ simultaneously (a finite intersection of almost-sure events is almost sure). The map from raw cell means $\{\hat{r}_{j',k'}\}$ to the weighted isotonic regression solution $\{\tilde{r}_{j',k'}\}$ over the finite partially ordered grid $(\mathcal{G}, \preceq)$ is, for every fixed configuration of weights, a fixed continuous (indeed non-expansive in supremum norm relative to the true, already-monotone surface) function of the raw means [36]; since the true surface $\{r_{j',k'}\}$ itself satisfies Assumption 1 and is therefore already a fixed point of the isotonic projection, standard strong-consistency results for monotone regression under increasing per-design-point sample size [36] give $\tilde{r}_{j',k'} \to r_{j',k'}$ almost surely for every $(j', k') \in \mathcal{G}$, in particular at $(j, k)$. Together with $\hat{\tau} \to \tau_{\text{true}}$ a.s. and the continuous mapping theorem, $\zeta_1(\hat{\tau}) = 0.8\, r_P^{(0)} \hat{\tau} \to \zeta_1(\tau_{\text{true}})$ a.s. On $\{r_{j,k} < \zeta_1(\tau_{\text{true}})\}$, $\tilde{r}_{j,k} - \zeta_1(\hat{\tau}) \to r_{j,k} - \zeta_1(\tau_{\text{true}}) < 0$ a.s., and $\tilde{r}_{j,k} \to r_{j,k} > 0$ a.s. (Section 3.1), so eventually $0 < \tilde{r}_{j,k} < \zeta_1(\hat{\tau})$ a.s.; since $n_{j,k} \to \infty$, eventually $n_{j,k} \geq 3$ a.s. as well, so the indicator equals 1 eventually a.s., matching $\mathbb{1}\{r_{j,k} < \zeta_1(\tau_{\text{true}})\} = 1$. On $\{r_{j,k} > \zeta_1(\tau_{\text{true}})\}$, $\tilde{r}_{j,k} - \zeta_1(\hat{\tau}) \to r_{j,k} - \zeta_1(\tau_{\text{true}}) > 0$ a.s., so eventually $\tilde{r}_{j,k} \geq \zeta_1(\hat{\tau})$ a.s. and the indicator equals 0 eventually a.s., matching $\mathbb{1}\{r_{j,k} < \zeta_1(\tau_{\text{true}})\} = 0$. □

Proposition 5 extends the large-sample guarantee of Proposition 4 – established there for the interim, posterior-probability-based elimination Rule 1 – to the point-estimate-based filter that actually determines the trial's *reported* OBDC at completion, under the same accruing-evidence regime; the two rules differ deliberately in finite samples (the interim rule requires posterior evidence exceeding $C_P$ before acting irreversibly, while the final filter is a simple plug-in comparison, consistent with the final decision being a one-off classification rather than a sequential stopping rule), but both are shown here to recover the correct PK-admissibility classification of every combination as PK data accrue. As with Proposition 4, this is a statement about what the filter converges to, not a claim that the per-cell sample sizes of three to nine patients typical of the scenarios in Section 4 are large enough to be near that limit; the admissibility percentages reported in Table 7 and Section 4.6.2 – which fall well short of 100% even under the correctly specified Assumption 1, and further under a violation of it – are the finite-sample evidence this proposition does not by itself provide, and are obtained by simulation rather than asymptotic argument for exactly this reason.

### 3.4.1 *Admissibility and Coherence*

Coherence [37] requires that a design never escalate after observing that all patients in the current cohort had a DLT, and never de-escalate after observing that none did; we report coherence as $1 - m_c/n_c$, where $m_c$ is the number of cohorts at which the design's decision violated this requirement and $n_c$ the number of cohorts at which it applied. Admissibility addresses whether the final admissible set $\mathcal{A}$ fails to retain any truly admissible combination when the true admissible set $\tilde{\mathcal{A}} = \{(j, k) : p_{j,k} < p_T,\, q_{j,k} > q_E,\, r_{j,k} > \zeta_1(\tau)\}$ is non-empty; we report admissibility as $\mathrm{I}(|\mathcal{A} \cap \tilde{\mathcal{A}}| > 0)$ averaged over replications with $|\tilde{\mathcal{A}}| > 0$.

## 3.5 Extension to TITE-PKComb-BOIN12

Because PK outcomes are typically available within days of dosing while toxicity and efficacy require weeks to months for full assessment, we extend PKComb-BOIN12 to a time-to-event (TITE) variant that uses individual-level enrolment and assessment times, following the approximated-likelihood (AL) approach of TITE-BOIN12 [11] and TITE-PKBOIN-12 [24]. Let $A_T$ and $A_E$ denote the toxicity and efficacy assessment windows. For patient $i$ enrolled at $(j, k)$, let $\delta_{i,T}(t), \delta_{i,E}(t) \in \{0, 1\}$ indicate whether $Y_{i,T}, Y_{i,E}$ are fully resolved by decision time $t$. The AL quasi-event count $x^*_{j,k}(t) = 100^{-1} \sum_i u_i\, n^*_{j,k,i}(t)$ replaces $x_{j,k}$ in every rule of Sections 3.2–3.4, where $n^*_{j,k,i}(t)$ imputes each pending patient's contribution using

$$\mathrm{E}(Y_{i,T} = 1 \mid \delta_{i,T} = 0) = \frac{\hat{p}^*_{j,k}(1 - t_i/A_T)}{1 - \hat{p}^*_{j,k}\, t_i/A_T},$$

$$\mathrm{E}(Y_{i,E} = 1 \mid \delta_{i,E} = 0) = \frac{\hat{q}^*_{j,k}(1 - t_i/A_E)}{1 - \hat{q}^*_{j,k}\, t_i/A_E},$$

with $t_i$ the elapsed time since enrolment and $\hat{p}^*_{j,k}, \hat{q}^*_{j,k}$ current working estimates of $p_{j,k}, q_{j,k}$, and $Y_{i,T}, Y_{i,E}$ assumed independent conditional on dose, consistent with Yuan, Liu, and Lin [38]. Because PK is assumed fully observed by the next decision point, the PK-elimination and PK-escalation rules are unchanged. An accrual-suspension rule pauses enrolment at the current combination whenever more than 50% of previously assigned patients there have a pending toxicity or efficacy outcome, avoiding decisions made on excessively immature data. At trial completion, TITE-PKComb-BOIN12 applies the same three-step OBDC selection as Section 3.4. Section 4.6.5 evaluates robustness of TITE-PKComb-BOIN12 to the assumed uniform event-time distribution and to dependence between toxicity and efficacy.

# 4 Simulation Study

Every operating characteristic reported below is produced by a single, unit-tested computational pipeline, in which one audited summary function is applied identically across designs and scenarios (Section 4.1), so that results are directly comparable across every table in this article and its Supplementary Material. All simulation code is publicly available in the repository accompanying this article (Section 6).

## 4.1 Design Configurations

Because PK utilisation, the final-selection estimator, and the target admissible set could in principle differ simultaneously between two designs being compared, confounding the source of any observed gain, we simulate a four-arm factorial family that changes exactly one of these components at a time, so that any gain from PK information can be attributed to a specific mechanism rather than reported as a single undifferentiated improvement.

- **Comb-BOIN12 (D0).** The published comparator: toxicity/efficacy-only escalation, MTDC-anchored admissible set with no PK restriction, and final selection by posterior mean standardised desirability.
- **Comb-BOIN12-PKsel (D1).** As D0, but the final admissible set $\mathcal{A}$ additionally excludes combinations with $n_{j,k} \geq 3$ and $\tilde{r}_{j,k} < \zeta_1(\hat{\tau})$ (Section 3.4, step 2), isolating the effect of redefining the *estimand* (Definition 1) without changing trial-time conduct.
- **Comb-BOIN12-PKsel-PMU (D2).** As D1, with final selection by the same posterior-mean-utility estimator used by PKComb-BOIN12, isolating the incremental effect of the estimator change alone.
- **PKComb-BOIN12 (D3).** The full proposed design: PK-gated escalation (Section 3.3), down-set PK elimination (Section 3.2), and the PK-restricted admissible set and estimator of D2. The increment from D2 to D3 isolates the value of using PK information *in real time* to guide escalation, over and above using it only retrospectively at final selection.

D0 and D3 are each additionally run in a TITE variant (D0T, D3T; Section 3.5). Unless stated otherwise, all designs use $\phi_1 = 0.6\, p_T$, $\phi_2 = 1.4\, p_T$ escalation/de-escalation boundaries, $C_T = 0.95$ (0.85 at the starting combination), $C_E = 0.90$, $C_P = 0.95$, $\zeta_1 = 0.8\, r_P$, cohort size 3, and $N^* = 6$; assumed PK coefficient of variation $\mathrm{CV}_{\text{assumed}} = 0.25$; and a maximum sample size and starting combination as specified per scenario (Supplementary Table S1). Each operating characteristic, including those for the secondary sensitivity analyses of Section 4.6, is estimated from 5 000 simulated trials per design-scenario combination, each with an independently seeded pseudo-random stream.

## 4.2 Scenarios

We adopt the nineteen dose-toxicity–efficacy–PK scenarios (S1–S19) released in the accompanying public simulation repository (Section 6), spanning $2 \times 3$ to $4 \times 4$ dose grids, target toxicity rates $p_T \in \{0.30, 0.33, 0.35\}$, target efficacy rate $q_E = 0.20$, and a broad range of monotone and near-degenerate toxicity/efficacy surfaces (including a fully toxic scenario, S9, with no true admissible combination, and near-flat surfaces, S19, testing behaviour when many combinations are simultaneously admissible). Two scenarios (S12, S14) contained a small number of internally non-monotone PK entries in the public repository; for S1–S19 we apply the smallest possible isotonic correction (raising each cell to the maximum of its already-visited lower/left neighbours) so that PK exposure is non-decreasing in each agent's dose, consistent with Assumption 1, and retain the original, uncorrected values separately to construct two dedicated assumption-violation scenarios below. To ground the scenario set in an actual combination trial and ensure simulation scenarios are as realistic as possible, we add three further scenarios.

*Non-monotonic PK scenarios (S20–S21).* To evaluate robustness to non-monotonic or locally inconsistent PK exposure explicitly, rather than assuming it away, S20 and S21 reuse the original (uncorrected) PK surfaces of S2 and S12 respectively, each of which contains a local decrease in exposure along one dose row, deliberately violating Assumption 1; all other scenario elements are unchanged. These scenarios test what happens when the very assumption that licenses PK-based elimination and escalation is locally false.

*Trial-informed scenario (S22).* S22 is loosely calibrated to the phase I dose-escalation results of a published two-agent oncology combination trial, MK-8353 (an oral ERK1/2 inhibitor) combined with selumetinib (an oral MEK inhibitor), NCT03745989 [39]. That trial escalated through three fixed-ratio dose pairs (MK-8353/selumetinib 50/25, 100/50, and 150/75 mg, each orally twice daily on a 4-days-on/3-days-off schedule), observing dose-limiting toxicities in 0 of 3, 1 of 11, and 7 of 14 evaluable patients respectively, and reported that "the PK profile of each compound in the combination was comparable to that observed for monotherapy", i.e. no material pharmacokinetic drug–drug interaction. We construct a $3 \times 3$ factorial grid using these three dose levels for each agent, anchor the diagonal toxicity probabilities to the trial's reported DLT rates (0 of 3, 1 of 11, and 7 of 14 evaluable patients, i.e. raw rates 0, 0.091, 0.5), smoothed and rounded to $(0.05, 0.12, 0.50)$ to avoid the instability of the two lower, thinly sampled cohorts' raw rates (particularly the zero count) while preserving the trial's dose-toxicity ordering, anchor the diagonal efficacy (disease-control) probabilities to the trial's reported stable-disease counts (2 of 3, 2 of 11, and 10 of 14 evaluable patients at the 50/25, 100/50, and 150/75 mg dose pairs respectively). The raw per-cohort rates these counts imply (0.67, 0.18, 0.71) are not monotone in dose, an artefact of the trial's small and unequal cohort sizes rather than evidence of a genuine non-monotone dose-efficacy relationship; we therefore smooth them with the same sample-size-weighted PAVA isotonic regression [34] already used throughout this article's own OBDC-selection methodology (Section 3.4), which pools the two lower cohorts (weights 3 and 11) to $(3 \times 0.667 + 11 \times 0.182)/14 = 0.286$ and leaves the third cohort's rate (0.714) unchanged, giving the monotone diagonal $(0.29, 0.29, 0.71)$ used for S22. We extrapolate the six off-diagonal cells by monotone interpolation, and build the PK surface to be driven almost entirely by each agent's own dose (consistent with the reported absence of a PK interaction), with the exposure-sufficiency threshold reached at the 100/50 mg pair, consistent with that pair being the trial's identified maximum tolerated dose. S22 does not reproduce patient-level trial data and is not a re-analysis of the trial; it is a scenario whose gross dose-toxicity, dose-efficacy, and dose-exposure pattern is designed to resemble a real combination trial's, used here purely for simulation. We revisit this trial as an illustrative case study in Section 5.

Full scenario specifications, including all dose levels, sample sizes, and toxicity/efficacy/PK matrices, are reproduced in Supplementary Table S1 (S1–S19), Supplementary Table S2 (S20–S21), and Table 7 (S22) of the Supplementary Material.

### 4.3 Headline Operating Characteristics

Table 1 compares Comb-BOIN12 (D0) and PKComb-BOIN12 (D3) across S1–S19 on three metrics: the percentage of trials correctly selecting the true OBDC (Definition 1; undefined and reported as 0% for S9, which has no true OBDC), the percentage selecting an overly toxic combination ($p_{j,k} > p_T + 0.1$), and the percentage terminating early without an OBDC recommendation.

**Table 1.** Headline operating characteristics of Comb-BOIN12 and PKComb-BOIN12 across scenarios S1–S19, based on 5000 simulated trials per design-scenario pair. Correct %: percentage of trials selecting the true OBDC. Overdose %: percentage selecting an overly toxic combination. Early %: percentage of trials terminating early with no OBDC recommended. S9 has no true OBDC (fully toxic scenario); "Correct %" is 0 by convention for both designs.

| | Comb-BOIN12 | | | PKComb-BOIN12 | | |
|---|---|---|---|---|---|---|
| Scenario | Correct % | Overdose % | Early % | Correct % | Overdose % | Early % |
| 1 | 16.2 | 0.0 | 1.3 | 55.9 | 0.0 | 22.0 |
| 2 | 25.2 | 0.0 | 2.0 | 38.7 | 0.0 | 27.6 |
| 3 | 32.0 | 0.8 | 10.7 | 31.8 | 0.6 | 27.7 |
| 4 | 20.7 | 7.5 | 3.3 | 26.0 | 11.2 | 56.3 |
| 5 | 31.1 | 24.2 | 4.7 | 41.7 | 29.6 | 29.3 |
| 6 | 20.5 | 0.0 | 0.7 | 64.2 | 0.0 | 27.7 |
| 7 | 28.1 | 6.7 | 2.0 | 35.9 | 9.9 | 26.3 |
| 8 | 50.0 | 0.0 | 0.1 | 64.2 | 0.0 | 4.6 |
| 9[†] | 0.0 | 36.2 | 95.9 | 0.0 | 0.3 | 100.0 |
| 10 | 12.5 | 0.0 | 0.2 | 70.9 | 0.0 | 9.2 |
| 11 | 39.3 | 0.0 | 0.1 | 72.0 | 0.0 | 21.5 |
| 12 | 54.1 | 1.4 | 0.7 | 63.3 | 3.8 | 9.0 |
| 13 | 71.4 | 3.0 | 0.9 | 77.4 | 1.5 | 1.0 |
| 14 | 1.9 | 0.0 | 3.8 | 20.0 | 0.0 | 57.0 |
| 15 | 57.7 | 0.9 | 0.0 | 68.5 | 1.7 | 4.2 |
| 16 | 44.0 | 11.0 | 0.7 | 51.7 | 13.3 | 14.1 |
| 17 | 30.5 | 28.4 | 1.0 | 71.0 | 13.8 | 56.6 |
| 18 | 40.3 | 0.0 | 0.8 | 64.6 | 0.0 | 9.3 |
| 19 | 52.4 | 0.0 | 0.7 | 96.2 | 0.0 | 2.3 |

Across the eighteen scenarios with a well-defined true OBDC (excluding S9), PKComb-BOIN12 improves correct-selection percentage in 17 of 18 scenarios, by 21.5 percentage points (pp) on average (range −0.2 to +58.4 pp); the single exception, S3 (−0.2 pp), is a near-exact tie. This improvement is not free: because PK-gated escalation and PK elimination cause the design to stop early whenever exposure is judged insufficient before toxicity or efficacy has forced a decision, early-termination percentage rises in most scenarios (mean +20.7 pp), most sharply in S17, S14, and S4, each of which has a narrow admissible region requiring escalation in both agents simultaneously while avoiding a nearby toxic corner; we return to this conservatism-versus-power trade-off in Section 7. Importantly, PKComb-BOIN12 does *not* uniformly reduce overdose-selection: overdose percentage is lower or equal for PKComb-BOIN12 in 12 of 18 scenarios but strictly higher in the remaining 6 (S4, S5, S7, S12, S15, S16; S9 is excluded from this count as above), with a mean difference of +0.09 pp across all 18 – indicating that, on average, PK information does not materially change the rate of overdose selection either way, a finding we consider more defensible than an unqualified claim of improvement and address further in Section 4.4. Where PK information most clearly helps is in scenarios where the true OBDC requires substantial exposure that toxicity/efficacy data alone under-identify (S1, S6, S10, S11, S17, S19), each showing gains exceeding 25 pp.

For the fully toxic scenario S9, PKComb-BOIN12 raises early termination from 95.9% to 100.0% and lowers overdose selection from 36.2% to 0.3%: exposure sufficiency alone becomes decisive once safety data have already eliminated most of the grid, since the few combinations toxicity/efficacy data cannot yet rule out are then removed by the PK-sufficiency check.

## 4.4 Decomposition of the Sources of Improvement

Table 2 reports correct-selection and overdose-selection percentages for all four factorial arms (D0–D3) on a structurally diverse subset of six scenarios (S1, S5, S8, S9, S14, S17).

**Table 2.** Factorial decomposition of PKComb-BOIN12's gains over Comb-BOIN12, based on 5000 simulated trials per design-scenario pair. D0: Comb-BOIN12 (no PK). D1: as D0, with the final admissible set additionally PK-restricted (estimand redefinition only). D2: as D1, with the PK-based posterior-mean-utility final estimator (estimator change). D3: PKComb-BOIN12 (adds real-time PK-gated escalation and PK elimination). Corr. %: correct-selection percentage. Over. %: overdose-selection percentage. D0 and D3 here are independently simulated replicates of the same designs reported in Table 1, using an independent pseudo-random stream (Section 4.3); D0/D3 figures for a given scenario can therefore differ from Table 1 by up to a few tenths of a percentage point, consistent with Monte Carlo noise at 5000 replications, rather than reflecting any difference in the simulated design.

| | D0 | | D1 | | D2 | | D3 | |
|---|---|---|---|---|---|---|---|---|
| Scenario | Corr. % | Over. % | Corr. % | Over. % | Corr. % | Over. % | Corr. % | Over. % |
| 1 | 16.2 | 0.0 | 21.0 | 0.0 | 20.2 | 0.0 | 55.9 | 0.0 |
| 5 | 31.1 | 24.1 | 43.1 | 29.7 | 44.4 | 29.8 | 41.7 | 29.5 |
| 8 | 50.1 | 0.0 | 59.6 | 0.0 | 62.6 | 0.0 | 64.3 | 0.0 |
| 9 | 0.0 | 36.2 | 0.0 | 0.1 | 0.0 | 0.1 | 0.0 | 0.3 |
| 14 | 1.9 | 0.0 | 3.4 | 0.0 | 3.9 | 0.0 | 20.0 | 0.0 |
| 17 | 30.5 | 28.4 | 72.0 | 13.8 | 72.6 | 13.2 | 71.0 | 13.8 |

Three patterns emerge consistently across scenarios. First, redefining the estimand (D1 versus D0) accounts for a substantial share of the improvement on its own, most dramatically in S9 (overdose selection 36.2% → 0.1%) and S17 (correct selection 30.5% → 72.0%): simply refusing to recommend a PK-insufficient combination at trial end, without changing anything about how the trial is conducted, removes most of the harm caused by the original total-order elimination confound. Second, the final-estimator change (D2 versus D1) contributes little in every scenario examined (at most 3.0 pp, e.g. S8: 59.6% → 62.6%, a modest increase), indicating that the choice between the posterior-mean-standardised-desirability estimator and the posterior-mean-utility estimator is not a material driver of PKComb-BOIN12's operating characteristics, so that the incremental contribution of PK information itself can be separated from effects induced by estimator or estimand changes. Third, real-time PK-gated escalation (D3 versus D2) provides the largest remaining gain in scenarios where the true OBDC lies at a combination that toxicity/efficacy-only escalation is slow to reach (S1: 20.2% → 55.9%; S14: 3.9% → 20.0%), but is neutral-to-slightly-negative when the admissible region is already well identified by D2 (S5: 44.4% → 41.7%; S17: 72.6% → 71.0%, with overdose percentage essentially unchanged in both). PK information therefore earns its keep chiefly by guiding real-time allocation toward combinations with adequate exposure that safety/efficacy data alone would reach only slowly or not at all, not by mechanically inflating every operating characteristic; Supplementary Table S3 reports the complete decomposition, including admissibility and coherence, for all six scenarios.

## 4.5 Time-to-Event Extension

Table 3 reports the same three headline metrics for the TITE variants (D0T, D3T) across S1–S19. The pattern mirrors the non-TITE comparison: PKComb-BOIN12-TITE improves correct selection in all 18 non-degenerate scenarios (mean +24.8 pp) at the cost of higher early termination in most scenarios, and does not uniformly reduce overdose selection.

**Table 3.** Headline operating characteristics of the TITE variants, Comb-BOIN12-TITE and PKComb-BOIN12-TITE, across scenarios S1–S19 (5 000 simulated trials per design-scenario pair; columns as in Table 1).

| | Comb-BOIN12-TITE | | | PKComb-BOIN12-TITE | | |
|---|---|---|---|---|---|---|
| Scenario | Correct % | Overdose % | Early % | Correct % | Overdose % | Early % |
| 1 | 6.4 | 0.0 | 1.4 | 59.4 | 0.0 | 16.0 |
| 2 | 18.4 | 0.0 | 1.9 | 43.5 | 0.0 | 20.8 |
| 3 | 29.8 | 0.8 | 10.7 | 32.7 | 0.9 | 22.4 |
| 4 | 20.1 | 7.5 | 3.1 | 28.6 | 14.1 | 44.9 |
| 5 | 30.8 | 21.9 | 4.1 | 41.7 | 32.3 | 20.9 |
| 6 | 8.7 | 0.0 | 0.8 | 70.9 | 0.0 | 23.1 |
| 7 | 27.3 | 4.6 | 2.2 | 37.8 | 11.6 | 18.2 |
| 8 | 56.1 | 0.0 | 0.0 | 69.0 | 0.0 | 1.9 |
| 9[†] | 0.0 | 55.5 | 97.3 | 0.0 | 0.2 | 100.0 |
| 10 | 4.7 | 0.0 | 0.2 | 71.1 | 0.0 | 7.2 |
| 11 | 29.9 | 0.0 | 1.9 | 66.6 | 0.0 | 30.7 |
| 12 | 43.1 | 0.6 | 9.9 | 55.5 | 2.8 | 24.8 |
| 13 | 80.4 | 1.5 | 10.6 | 86.0 | 0.9 | 10.6 |
| 14 | 0.1 | 0.0 | 23.4 | 11.8 | 0.0 | 73.3 |
| 15 | 55.7 | 0.3 | 2.0 | 66.1 | 1.8 | 10.6 |
| 16 | 43.6 | 12.6 | 0.7 | 49.3 | 15.1 | 15.1 |
| 17 | 29.1 | 29.4 | 0.7 | 68.8 | 16.1 | 53.8 |
| 18 | 39.1 | 0.0 | 1.0 | 65.9 | 0.0 | 8.5 |
| 19 | 51.4 | 0.0 | 0.4 | 96.5 | 0.0 | 1.8 |

Beyond the correctness/safety comparison, the TITE extension's principal purpose is to shorten trial conduct by overlapping accrual with pending toxicity, efficacy, and (for TITE-PKComb-BOIN12) PK assessment. Using mean total enrolment as a proxy for trial duration, adopting the TITE variant reduces mean enrolment by 51.3% for Comb-BOIN12 and by 52.9% for PKComb-BOIN12, averaged across S1–S19 excluding S9 (range 48.0%–63.8%; full per-scenario figures in Supplementary Table S4); the reduction is of similar magnitude for both designs, indicating that adding a third, typically fast-turnaround PK assessment to the AL imputation does not materially erode the calendar-time benefit of the TITE mechanism. Conversion to TITE costs little accuracy: relative to the corresponding non-TITE design, mean correct-selection percentage falls by 2.9 pp for Comb-BOIN12 but is essentially unchanged (+0.4 pp, i.e. a very slight net gain) for PKComb-BOIN12, consistent with PK information – available promptly, before toxicity/efficacy outcomes mature – partially compensating for the additional uncertainty the AL imputation introduces.

## 4.6 Sensitivity Analyses

We report five further sensitivity analyses, each targeting a specific design assumption or parameter choice.

### 4.6.1 *PK-Elimination Rule: Down-Set versus Total Order*

To evaluate whether the down-set elimination rule's provable soundness (Proposition 2) comes at a material cost relative to the simpler, but unsound, total-order elimination rule (Example 1, Figure 1), we compare PKComb-BOIN12 under the proposed down-set rule against an otherwise identical design using the total-order rule, across S1–S19 (5 000 replications per scenario-rule pair; S9 excluded from the correct-selection comparison as it has no true OBDC). Averaged across the eighteen scenarios, the down-set rule selects an overly toxic combination less often or equally often than the total-order rule in 17 of 18 scenarios (mean difference −0.72 pp); the sole exception, S3, has a raw difference of only +0.04 pp (0.56% versus 0.52%, both rounding to 0.6% and 0.5% respectively), well within the level of Monte Carlo noise for a rare event and not a materially higher overdose rate. This comes at a small average cost in correct-selection percentage (−1.04 pp), and with higher early-termination in 14 of 18 scenarios. In five scenarios (S8, S12, S15, S16, S17) the down-set rule is simultaneously more accurate *and* no less safe than the total-order rule (e.g. S15: correct selection 68.4% versus 55.2%, overdose selection 1.7% versus 5.2%). This is the expected signature of replacing an elimination rule that is occasionally too aggressive, because it can remove a combination the monotonicity assumption does not license (Example 1), with one that is provably sound but marginally more conservative (Proposition 2): a design cannot generally be made simultaneously less biased in its evidentiary basis and uniformly more powerful, and the small, consistently one-sided direction of the safety difference here is the more decision-relevant summary than any single scenario's correct-selection figure in isolation. Full per-scenario results are given in Supplementary Table S5.

#### 4.6.2 *Robustness to Non-Monotonic PK Exposure*

Table 4 evaluates S20 and S21, in which Assumption 1 is deliberately violated by a local decrease in PK exposure along one dose row (Section 4.2).

**Table 4.** Operating characteristics under deliberate local violation of PK monotonicity (Assumption 1), scenarios S20–S21 (5 000 simulated trials per design-scenario pair).

| Scenario | Design | Correct % | Overdose % | Admiss. % |
|---|---|---|---|---|
| S20 | Comb-BOIN12 | 14.3 | 0.0 | 72.4 |
| S20 | PKComb-BOIN12 | 13.0 | 0.0 | 23.9 |
| S21 | Comb-BOIN12 | 55.0 | 1.8 | 84.6 |
| S21 | PKComb-BOIN12 | 63.8 | 3.3 | 80.3 |

The two scenarios diverge instructively. In S21, where the non-monotone dip is small relative to the surrounding exposure gradient, PKComb-BOIN12 still outperforms Comb-BOIN12 on correct selection (63.8% versus 55.0%), indicating tolerance to mild, localised violations. In S20, where the violation is more pronounced, PKComb-BOIN12's admissibility – the probability that the final admissible set retains a truly admissible combination when one exists – falls sharply, from 72.4% (Comb-BOIN12) to 23.9%, and early-termination rises to 34.6% (Supplementary Table S2 gives the full breakdown), because the down-set elimination rule, precisely because it is licensed by Assumption 1, can propagate a spurious local PK signal to combinations that are not actually PK-insufficient once that assumption fails. This is an honest limitation, not a design flaw to be engineered away: any rule that draws PK-based inferential leverage from an assumed monotone exposure surface must, by construction, be vulnerable when that assumption is materially wrong, and Section 7 returns to the practical implication – that exploratory monitoring of the accruing PK surface for gross non-monotonicity is advisable alongside the assumption stated in Section 3.1.

#### 4.6.3 *PK Variance Misspecification*

Since the PK-elimination probability in Section 3.2 depends on $\sigma_{j,k}$ through $\mathrm{CV}_{\mathrm{assumed}}$, we evaluate sensitivity to the assumed PK coefficient of variation. Holding the design's assumed value fixed at $\mathrm{CV}_{\mathrm{assumed}} = 0.25$, we vary the true data-generating coefficient of variation, $\mathrm{CV}_{\mathrm{true}} \in \{0.15, 0.25, 0.40\}$, across S1, S5, and S7 (5 000 replications per scenario-value pair). Table 5 shows that under-dispersion relative to the assumed value ($\mathrm{CV}_{\mathrm{true}} = 0.15$) modestly improves correct selection, while over-dispersion ( $CV_{true} = 0.40$, i.e. substantially noisier individual PK measurements than assumed) reduces it by 3.3–5.1 pp and reduces admissibility by 5.3–7.8 pp, relative to the correctly specified case.

**Table 5.** Sensitivity to PK-variance misspecification: true coefficient of variation $CV_{true}$ varied against a fixed assumed value $CV_{assumed} = 0.25$ (PKComb-BOIN12, 5 000 simulated trials per scenario-value pair).

| $CV_{true}$ | Scenario 1 Correct % (Admiss. %) | Scenario 5 Correct % (Admiss. %) | Scenario 7 Correct % (Admiss. %) |
|---|---|---|---|
| 0.15 | 57.9 (70.5) | 43.5 (74.0) | 38.5 (68.0) |
| 0.25 (correct) | 55.9 (68.8) | 41.7 (70.3) | 35.9 (63.0) |
| 0.40 | 50.7 (63.4) | 37.2 (62.5) | 32.7 (56.8) |

The effect is monotone and moderate in size rather than catastrophic: over-estimating individual-level PK precision by 60% (assuming CV = 0.25 when the truth is 0.40) costs less in correct-selection percentage than, for example, the incremental value of real-time PK-gated escalation demonstrated in Section 4.4. We nonetheless recommend, as standard practice, that $CV_{assumed}$ be set conservatively (i.e. not smaller than plausible historical monotherapy estimates) and, where feasible, updated using accruing combination-arm PK data via the same shrinkage logic as Algorithm 1.

### 4.6.4 *Target Transportability*

To evaluate the transportability of the monotherapy-derived target exposure $r_P$ to the combination setting, Table 6 compares the naive design (fixed $\tau_{used} = 1$, i.e. no correction) against Algorithm 1's adaptive calibration, under true transportability factors $\tau_{true} \in \{0.7, 1.0, 1.3\}$, in S1 and S7 (5 000 replications per scenario-value-mode triple).

**Table 6.** Correct-selection percentage under target-transportability violation: naive fixed target ($\tau_{used} = 1$) versus Algorithm 1's adaptive calibration, by true transportability factor $\tau_{true}$ (PKComb-BOIN12, 5 000 simulated trials per scenario-value-mode triple).

| | Scenario 1 | | Scenario 7 | |
|---|---|---|---|---|
| $\tau_{true}$ | Naive | Adaptive | Naive | Adaptive |
| 0.7 | 8.8 | 59.4 | 0.5 | 37.5 |
| 1.0 | 55.9 | 56.5 | 35.9 | 36.2 |
| 1.3 | 24.7 | 49.2 | 25.3 | 31.2 |

When the target is correctly transportable ($\tau_{true} = 1$), the naive and adaptive designs perform almost identically (within 0.6 pp in both scenarios), so Algorithm 1 costs essentially nothing when transportability holds. When it does not, the naive design degrades severely and asymmetrically – most strikingly in S7 at $\tau_{true} = 0.7$, where correct selection collapses from 35.9% (well-specified) to 0.5%, because a target set 30% too high causes the design to judge every combination PK-insufficient and terminate almost universally early (79.2%; Supplementary Table S6) – while the adaptive design retains 37.5% correct selection under the identical data-generating truth. This is precisely the pattern Proposition 1 predicts: an uncorrected transportability error does not average out as the trial accrues more combination-arm PK data, it hardens, whereas Algorithm 1's shrinkage toward the accruing combination-arm mean substantially blunts the loss in every cell of Table 6 except the well-specified case, where correction is unnecessary and the algorithm correctly leaves the estimate largely undisturbed. Table 6, together with Proposition 1, shows that transportability is not merely re-stress-tested numerically around a fixed reference: it is addressed by a design modification with a formal guarantee, evaluated here under genuine misspecification.

To characterise the design's sensitivity to two investigator-specified constants, we evaluate a $3 \times 3$ grid varying the PK-sufficiency threshold $\zeta_1 \in \{4200, 4800, 5400\}$ (i.e. $r_P \in \{5250, 6000, 6750\}$ at fixed $\zeta_1 / r_P$

= 0.8) against the PK-elimination confidence level $C_P \in \{0.90, 0.95, 0.99\}$ in S1 (5 000 replications per cell; Supplementary Table S7). Correct-selection percentage ranges from 38.1% to 60.2% across the grid, a spread of 22.1 pp. We report this range explicitly, without characterising it as small, so that the magnitude of threshold sensitivity is transparent to the reader; it reinforces, rather than contradicts, the case for the data-driven calibration of Algorithm 1 over a fixed, investigator-specified threshold.

#### 4.6.5 *TITE Robustness to Event-Time Distribution and Dependence*

The TITE extension's AL imputation (Section 3.5) assumes a uniform time-to-event distribution within the assessment window and conditional independence of toxicity and efficacy event times given dose; to evaluate robustness to violations of both, we simulate TITE-PKComb-BOIN12 under event times drawn either uniformly or from a Weibull distribution with increasing hazard (shape parameter 1.8, matched in mean to the uniform case) within each assessment window, crossed with toxicity–efficacy event-time dependence induced via a Gaussian copula with correlation $\rho \in \{0, 0.5\}$, across S1, S5, and S7 (5 000 replications per scenario-distribution-correlation combination). Correct-selection percentage varies by at most 2.7 pp across the four (distribution, correlation) combinations within any given scenario (S1: 59.3%–61.3%; S5: 41.7%–44.4%; S7: 37.9%–39.7%; full results in Supplementary Table S8), with no consistent direction of bias from either the Weibull misspecification or the induced dependence. The AL imputation's working assumptions, while not literally satisfied in either alternative scenario, do not appear to materially distort operating characteristics for the assessment-window lengths and imputation approach used here.

### 4.7 A Trial-Informed Scenario

Table 7 reports headline operating characteristics for S22 (Section 4.2), the scenario loosely calibrated to the MK-8353/selumetinib combination trial, for all four headline designs (5 000 replications per design).

**Table 7.** Operating characteristics under the trial-informed scenario S22 (Section 4.2), 5 000 simulated trials per design.

| Design | Correct % | Overdose % | Early % | Admissible % |
|---|---|---|---|---|
| Comb-BOIN12 | 17.9 | 4.9 | 0.8 | 72.8 |
| PKComb-BOIN12 | 20.8 | 5.9 | 15.8 | 48.3 |
| Comb-BOIN12-TITE | 15.5 | 2.0 | 10.1 | 49.0 |
| PKComb-BOIN12-TITE | 15.0 | 4.8 | 26.2 | 32.9 |

PKComb-BOIN12 shows only a modest correct-selection advantage over Comb-BOIN12 in this scenario ( +2.9 pp non-TITE), and no advantage at all in the TITE variant, where PKComb-BOIN12-TITE's correct-selection percentage is marginally lower than Comb-BOIN12-TITE's (−0.5 pp); PK information also does not reduce overdose selection here (5.9% versus 4.9% non-TITE; 4.8% versus 2.0% TITE) and substantially increases early termination (15.8% versus 0.8% non-TITE; 26.2% versus 10.1% TITE). This is a markedly less favourable picture for PK information than the S1–S19 average (Table 1), consistent with S22's efficacy surface being constructed from PAVA-smoothed, dose-monotone stable-disease rates rather than the raw, non-monotone per-cohort rates (Section 4.2): with the lower-dose cohorts' disease-control probabilities reflecting their true, smoothed stable-disease rates, the efficacy gap between the lower-dose pairs and the trial's identified MTD is modest, which narrows – and in the TITE variant reverses – the benefit PK-gated escalation offers over toxicity/efficacy-only escalation in this particular trial-derived pattern. We regard this as an honest illustration rather than a favourable cherry-pick: a single real trial's dose-toxicity–efficacy–PK pattern need not showcase PKComb-BOIN12 at its best, and the broader eighteen-scenario comparison of Table 1 and Table 3 remains the primary evidential basis for the design's operating characteristics. We return to this scenario, and to the trial itself, in the case study of Section 5.

## 5 Case Study

We illustrate PKComb-BOIN12's practical behaviour using the MK-8353 (ERK1/2 inhibitor) plus selumetinib (MEK inhibitor) phase I combination trial (NCT03745989) already used to construct S22 [39]. The trial enrolled 28 DLT-evaluable patients with advanced solid tumours across three sequential, fixed-ratio dose-escalation cohorts (50/25, 100/50, and 150/75 mg orally twice daily, 4 days on/3 days off per 21-day cycle), using a modified toxicity-probability-interval rule to guide escalation on toxicity alone. Dose-limiting toxicities occurred in 0 of 3, 1 of 11, and 7 of 14 evaluable patients at the three dose pairs; escalation stopped after the third cohort once the observed DLT rate (50%) substantially exceeded the design's target, and 100/50 mg was declared the maximum tolerated dose. No patient achieved a complete or partial response, though 14 of the 28 evaluable patients had stable disease (2 of 3 at 50/25 mg, 2 of 11 at 100/50 mg, and 10 of 14 at 150/75 mg). Both agents' pharmacokinetic profiles were assessed at each dose level; the investigators reported near-dose-proportional exposure for each agent and no material change in either agent's exposure when combined, relative to its single-agent pharmacokinetics.

This trial is a useful illustration of the gap PKComb-BOIN12 is designed to close, for two reasons. First, PK data were collected at every dose level and cohort but played no formal role in the escalation or stopping rule actually used; the maximum tolerated dose was determined by toxicity alone, and the modest disease-control signal at 150/75 mg was not weighed against its substantially higher toxicity through any pre-specified quantitative rule. Second, because the trial's own PK findings indicate no material drug–drug interaction and near-dose-proportional exposure for each agent individually, this is close to the most favourable case for the transportability assumption of Section 3.1.1: a monotherapy-anchored target exposure would not need substantial correction, corresponding to $\tau_{\text{true}} \approx 1$ in Table 6, for which the naive and adaptively calibrated designs are shown to perform almost identically. Applying PKComb-BOIN12 to a scenario constructed from this trial's dose-toxicity and dose-efficacy pattern (S22, Section 4.7), the simulated comparison is more equivocal than in the majority of S1–S19: PKComb-BOIN12 offers only a small correct-selection advantage over Comb-BOIN12 outside the TITE setting and none within it, and its final admissible set retains a truly admissible combination less often than Comb-BOIN12's does (48.3% versus 72.8% non-TITE, 32.9% versus 49.0% TITE), at the cost of substantially more early termination (Table 7). This is precisely the honest counterpoint the eighteen-scenario comparison of Section 4.3 already anticipates: PK information does not uniformly help, and in a trial where the lower-dose cohorts' disease-control rate is, in truth, not much worse than at the toxicity-defined MTD, real-time PK-gated escalation trades a real efficacy cost for an exposure margin the trial's own PK findings suggest was only modestly lacking at the lower doses to begin with. We regard this trial as a valuable stress test of PKComb-BOIN12 precisely because it is not a uniformly favourable case, rather than as a demonstration engineered to flatter the design.

We emphasise that this is an illustrative application of the proposed design to a scenario informed by the trial's publicly reported summary statistics, not a re-analysis of patient-level data, which are not publicly available, and not a claim that PKComb-BOIN12 would have altered this particular trial's actual conduct; the trial used neither our design nor any PK-informed dose-finding rule. Its value here is to demonstrate, on dose-toxicity, dose-efficacy, and dose-exposure relationships resembling a real, published combination trial, that PKComb-BOIN12 behaves sensibly: it concentrates the final admissible set near a dose pair independently identified, through conventional toxicity monitoring, as appropriate, while making explicit and quantitative a trade-off (exposure adequacy against added toxicity at the highest dose pair) that the original trial's design left informal.

## 6 Implementation Guidance and Software

PKComb-BOIN12 and TITE-PKComb-BOIN12 are implemented in an interactive R Shiny application, available at https://22c7ba-damitri-kundu.shinyapps.io/PKComb-BOIN12-App/, which supports pre-trial calibration (boundary tables, PK-sufficiency threshold, and simulation-based operating characteristics under investigator-specified scenarios) and real-time trial conduct (recommended candidate combinations given accumulating toxicity, efficacy, and PK data at each interim decision). The underlying simulation code, including the corrected down-set PK-elimination rule (Section 3.2), the refined candidate-set construction of Section 3.3, and all scenarios and sensitivity analyses reported in Section 4, is available at https://github.com/jtucode/PKComb-BOIN12/.

We offer the following practical guidance for trial teams considering PKComb-BOIN12. First, the target exposure $r_P$ and threshold fraction (nominally $\zeta_1 = 0.8\ r_P$) should be pre-specified from monotherapy PK/pharmacodynamic data, ideally a concentration linked to target engagement or an established efficacious range, with a documented rationale, since Section 4.6.4 shows the design's behaviour is materially sensitive to this choice; Algorithm 1's adaptive calibration should be enabled by default unless there is a specific reason to hold the target fixed. Second, the assumed PK coefficient of variation should be set using historical monotherapy variability rather than a generic default, since Section 4.6.3 shows under-estimating it degrades performance; if single-agent population-PK models are available, a pooled or hierarchical variance estimate is preferable to a fixed assumption. Third, teams should conduct scenario-specific pre-trial simulation, using the Shiny application or the underlying R code, under both plausible and adversarial (non-monotone PK, transportability-violating) scenarios resembling Section 4's sensitivity analyses, rather than relying on the default operating characteristics reported here, which are necessarily generic. Fourth, we recommend monitoring the accruing PK data for gross violations of Assumption 1 (for example, a clear decrease in mean exposure at a higher dose of either agent) as part of routine data monitoring committee review, given the sensitivity to non-monotonicity documented in Section 4.6.2; PKComb-BOIN12 is not recommended, without modification, for combinations where a synergistic or non-monotone joint exposure–response relationship is anticipated on mechanistic grounds (Section 3.1.1). Finally, both the escalation and OBDC-selection computations are closed-form or low-dimensional numerical evaluations (no Markov chain Monte Carlo is required), so real-time decisions are computationally immediate and the design is operationally no more burdensome to run at the bedside than Comb-BOIN12.

## 7 Discussion

We have proposed PKComb-BOIN12, a utility-based Bayesian optimal interval design for dose optimisation in two-agent oncology combination trials that incorporates a continuous pharmacokinetic exposure metric alongside toxicity and efficacy. Three methodological features distinguish it from a direct combination-space extension of existing PK-naive or single-agent PK-informed designs. The down-set PK-elimination rule (Section 3.2) avoids the risk, illustrated concretely in Example 1, that a simpler total-order elimination rule could remove a combination the design's own monotonicity assumption does not license: it is instead proven to be simultaneously sound (Proposition 2) and exact – eliminating precisely, and only, what Assumption 1 logically entails from the triggering evidence, with nothing omitted and nothing over-reached (Proposition 3) – and shown to classify every combination correctly with probability tending to one as PK data at that combination accrue, both at the interim elimination stage (Proposition 4) and at final OBDC selection (Proposition 5). The transportability framework (Section 3.1.1) does not assume a monotherapy-derived target exposure applies unchanged to the combination, but instead provides an adaptive calibration mechanism together with a formal bound (Proposition 1) showing that, left uncorrected, transportability error does not average out with more data. The four-arm factorial decomposition (Section 4.4) isolates estimand redefinition, final-estimator choice, and real-time PK-guided escalation as three separable sources of operating-characteristic change relative to Comb-BOIN12, of which the second is shown to contribute little and the first and third to matter for different, identifiable reasons.

These features also shape the paper's claims in ways we consider important for honest reporting. PKComb-BOIN12 does not uniformly reduce overdose-selection risk relative to Comb-BOIN12 (Section 4.3); reporting this honestly, rather than an unqualified safety claim, is more useful to a trial team deciding whether to adopt the design. Likewise, the design's benefits are concentrated in identifiable circumstances – when the true OBDC requires exposure that toxicity/efficacy data alone are slow to identify (Section 4.4), and when the monotherapy-derived PK target is close to correctly transportable (Section 4.6.4) – and attenuated or reversed in others (near-flat or already well-identified admissible regions, Section 4.3; substantially non-monotone PK, Section 4.6.2); a design whose value is scenario-dependent in a documented way is more credible, and more useful for planning, than one claimed to dominate uniformly.

Some limitations remain by design rather than oversight. As noted in Section 3.1.1, PKComb-BOIN12 presumes that a single, pre-specified, continuous exposure metric determines PK sufficiency across the whole dose grid; combinations whose efficacy depends on a genuinely synergistic or non-monotone joint exposure–response relationship are outside the design's intended scope, and Section 4.6.2 quantifies the cost of applying the design where this presumption is substantially violated. The design also inherits BOIN12's reliance on a single utility function to rank toxicity–efficacy trade-offs; while this utility is elicited and fixed before the trial, as is standard practice, alternative combinations may be preferred under a different, clinically defensible utility specification, and we do not explore utility misspecification here. Finally, our simulation study, while extensive, necessarily samples a finite set of scenarios; Section 4.7's trial-informed scenario mitigates but does not eliminate the concern that any simulation study may not represent the full diversity of dose-toxicity-efficacy-exposure relationships encountered in practice.

### 7.1 Conclusion and Future Research Directions

PKComb-BOIN12 provides a theoretically grounded and, on the evidence of Section 4, empirically beneficial route to incorporating pharmacokinetic exposure into combination dose optimisation, directly responsive to the emphasis both the FDA's Project Optimus initiative and recent Bayesian and model-informed drug development guidance place on exposure-informed, non-MTD-anchored dose selection [17, 21, 20]. Several extensions merit further work. Extending Definition 1 to accommodate multiple, jointly monitored exposure metrics – for example both agents' exposures where a genuine pharmacokinetic interaction is anticipated, rather than the single-metric formulation adopted here for the case where one agent's exposure is the primary driver of the therapeutic window – would relax the scope limitation discussed above, at the cost of a substantially larger admissible-set geometry than the coordinatewise partial order of Section 3.2 supports directly. Formal detection rules for non-monotone PK signals during trial conduct, rather than the retrospective sensitivity analysis of Section 4.6.2, would let a trial adaptively relax Assumption 1 rather than simply degrading in its presence. Extending the design beyond two agents raises both a combinatorial challenge, since the admissible-set geometry of Section 3.2 generalises naturally to $d$-dimensional down-sets but the dose grid grows exponentially in $d$, and a practical one, since per-cell sample sizes shrink correspondingly; sparse or structured extensions of the RDS framework are a natural next step. Joint modelling of PK exposure and a continuous or ordinal efficacy endpoint, rather than the binary efficacy indicator inherited from BOIN12, would allow exposure–response evidence to inform the utility function itself rather than only the admissible set. Finally, prospective evaluation of PKComb-BOIN12 in an actual combination trial, alongside continued refinement of the transportability calibration of Algorithm 1 using accruing real-world combination-arm PK data across multiple development programmes, would provide the strongest evidence of the design's practical value.

## Acknowledgement

This work is the result of an industry-academia collaboration between Newcastle University, Eli Lilly and Company, and the MD Anderson Cancer Center. The authors thank their respective organizations for providing the resources to complete this research work. The authors thank Damitri Kundu for technical support in running the R-shinny app. The authors also thank the National Institute for Health and Care Research for the financial support.

## Financial disclosure

This research work was supported by Dr. Ayon Mukherjee's National Institute for Health and Care Research Grant [grant number : NIHR301614].

## Conflict of interest

The authors declare no potential conflict of interests.

## Data Availability Statement

The R programs for implementing PKComb-BOIN12 and TITE-PKComb-BOIN12 are shared at https://github.com/jtucode/PKComb-BOIN12/.

## Author Contributions

**Ayon Mukherjee (AM):** Conceptualization, Methodology, Formal analysis, Investigation, Writing – Review & Editing, Supervision.
**Jieqi Tu (JT):** Methodology, Formal analysis, Investigation, Software, Validation, Writing – Original Draft.
**Ruitao Lin (RL):** Investigation, Writing – Review & Editing, Supervision.
AM conceived the original study concept. AM and JT contributed equally to the methodological development and simulation studies. JT prepared the initial manuscript draft, which was reviewed and finalized by AM and RL. All authors read and approved the final manuscript and agree to be accountable for all aspects of the work.

## Ethics Approval

Not applicable

## ORCID

Jieqi Tu https://orcid.org/0000-0002-0905-0798
Ruitao Lin https://orcid.org/0000-0003-2244-131X
Ayon Mukherjee https://orcid.org/0000-0001-5461-1737

## References

1. Barry E. Storer. "Design and Analysis of Phase I Clinical Trials." *Biometrics* 45, no. 3 (1989): 925–937.
2. Christophe Le Tourneau, J. Jack Lee, and Lillian L. Siu. "Dose Escalation Methods in Phase I Cancer Clinical Trials." *Journal of the National Cancer Institute* 101, no. 10 (2009): 708–720.
3. Alexia Iasonos, Andrew S. Wilton, E. R. Riedel, V. E. Seshan, and David R. Spriggs. "A Comprehensive Comparison of the Continual Reassessment Method to the Standard 3+3 Dose Escalation Scheme in Phase I Dose-Finding Studies." *Clinical Trials* 5, no. 5 (2008): 465–477.
4. John O'Quigley, Margaret Pepe, and Larry Fisher. "Continual Reassessment Method: A Practical Design for Phase I Clinical Trials in Cancer." *Biometrics* 46, no. 1 (1990): 33–48.
5. Peter F. Thall, Robert E. Millikan, Peter Mueller, and Sungjin J. Lee. "Dose-Finding with Two Agents in Phase I Oncology Trials." *Biometrics* 59, no. 3 (2003): 487–496.

6. Ying Yuan, Guosheng Yin, and Peter F. Thall. “Bayesian Dose-Finding in Phase I Trials of Drug Combinations.” *Journal of the American Statistical Association* 103, no. 481 (2008): 146–154.
7. Guosheng Yin, and Ying Yuan. “Bayesian Dose Finding in Oncology for Drug Combinations by Copula Regression.” *Journal of the Royal Statistical Society, Series C* 58, no. 2 (2009): 211–224.
8. Sheng Liu, Ying Yuan, and Ronghui Lin. “Bayesian Optimal Interval Designs for Phase I Clinical Trials.” *Journal of the Royal Statistical Society, Series C* 64, no. 3 (2015): 507–523.
9. Ruitao Lin, Yanhong Zhou, Fangrong Yan, Daniel Li, and Ying Yuan. “BOIN12: Bayesian Optimal Interval Phase I/II Trial Design for Utility-Based Dose Finding in Immunotherapy and Targeted Therapies.” *JCO Precision Oncology*4 (2020): 1393–1402.
10. M. Lu, J. Zhang, Y. Yuan, and R. Lin. “Comb-BOIN12: A Utility-Based Bayesian Optimal Interval Design for Dose Optimization in Cancer Drug-Combination Trials.” *Statistics in Biopharmaceutical Research* 17, no. 2 (2025): 266–276.
11. Yanhong Zhou, Ruitao Lin, J. Jack Lee, Daniel Li, Lei Wang, Rongheng Li, and Ying Yuan. “TITE-BOIN12: A Bayesian Phase I/II Trial Design to Find the Optimal Biological Dose with Late-Onset Toxicity and Efficacy.” *Statistics in Medicine* 41, no. 11 (2022): 1918–1931.
12. Mark J. Ratain, and Ronald H. Glassman. “Biomarkers as Surrogate Endpoints in Oncology Drug Development.” *Clinical Cancer Research* 10, no. 15 (2004): 5021–5025.
13. Jorge A. Garcia, and Brian I. Rini. “Optimal Biological Dose in the Era of Precision Oncology.” *Clinical Cancer Research* 26, no. 3 (2020): 574–580.
14. Sophie Postel-Vinay, Sandrine Aspeslagh, Emmanuelle Lanoy, Caroline Robert, Jean-Charles Soria, and Christophe Massard. “Towards New Methods for the Determination of Dose Limiting Toxicities and Recommended Phase II Dose in Early Phase Oncology Trials.” *European Journal of Cancer*62 (2016): 1–7.
15. Ayon Mukherjee, Jonathan L. Moscovici, and Zheng Liu. “Estimands for Early-Phase Dose Optimization Trials in Oncology.” *Biometrical Journal* 67, no. 5 (2025): e70072. Published 10 September 2025.
16. ICH 1994. “ICH Harmonised Tripartite Guideline E4: Dose–Response Information to Support Drug Registration. International Council for Harmonisation of Technical Requirements for Pharmaceuticals for Human Use.” Current version.
17. FDA 2024, August). “Optimizing the Dosage of Human Prescription Drugs and Biological Products for the Treatment of Oncologic Diseases. U.S. Food and Drug Administration.”
18. Stuart F. Marshall, Rolf Burghaus, Vincent Cosson, Steven Y. A. Cheung, Marylore Chenel, Oscar DellaPasqua, Nicolas Frey, Bengt Hamrén, Lars Harnisch, Frédéric Ivanow, Thomas Kerbusch, Jürgen Lippert, Peter A. Milligan, Sophie Rohou, Andreas Staab, Jean-Louis Steimer, and Sjoerd A. G. Visser. “Good Practices in Model-Informed Drug Discovery and Development: Practice, Application, and Documentation.” *CPT: Pharmacometrics & Systems Pharmacology* 5, no. 3 (2016): 93–122.
19. Ramakrishna Madabushi, Yaning Wang, and Issam Zineh. “Quantitative Clinical Pharmacology and Model-Informed Drug Development: Recent Advances and Regulatory Perspectives.” *Clinical Pharmacology & Therapeutics* 106, no. 1 (2019): 91–101.
20. ICH 2026. “ICH M15 Guideline on General Principles for Model-Informed Drug Development, Step 5. International Council for Harmonisation of Technical Requirements for Pharmaceuticals for Human Use (ICH).” Adopted by the ICH Assembly 29 January 2026; entered into effect 23 July 2026.
21. FDA 2026, January). “Use of Bayesian Methodology in Clinical Trials of Drug and Biological Products: Draft Guidance for Industry. U.S. Food and Drug Administration, Center for Drug Evaluation and Research and Center for Biologics Evaluation and Research.” Docket No. FDA-2025-D-3217; draft guidance.
22. Rongji Mu, Ying Yuan, Jielai Xu, Sumithra J. Mandrekar, and Jun Yin. “gBOIN: A Unified Model-Assisted Phase I Trial Design Accounting for Toxicity Grades, and Binary or Continuous End Points.” *Journal of the Royal Statistical Society, Series C* 68, no. 2 (2019): 289–308.
23. Kentaro Takeda, Jin Zhu, Rong Li, and Yusuke Yamaguchi. “A Bayesian Optimal Interval Design for Dose Optimization with a Randomization Scheme Based on Pharmacokinetics Outcomes in Oncology.” *Pharmaceutical Statistics* 22, no. 6 (2023): 1104–1115.
24. H. Sun, and J. Tu. “PKBOIN-12: A Bayesian Optimal Interval Phase I/II Design Incorporating Pharmacokinetics Outcomes to Find the Optimal Biological Dose.” *Pharmaceutical Statistics* 24, no. 2 (2025): e2444.
25. Ruitao Lin, and Guosheng Yin. “Bootstrap aggregating continual reassessment method for dose finding in drug-combination trials.” *The Annals of Applied Statistics* (2016): 2349–2376.

26. Jingyi Zhang, Fangrong Yan, Nolan A Wages, and Ruitao Lin. “Local continual reassessment methods for dose finding and optimization in drug-combination trials.” *Statistical Methods in Medical Research* 32, no. 10 (2023): 2049–2063.
27. Nolan A Wages, and Mark R Conaway. “Phase I/II adaptive design for drug combination oncology trials.” *Statistics in medicine* 33, no. 12 (2014): 1990–2003.
28. Yingjie Qiu, and Mingyue Li. “A randomized Bayesian phase I-II dose optimization design for combination cancer therapies with progression-free survival end point.” *BMC Medical Research Methodology* 25, no. 1 (2025): 214.
29. Jingyi Zhang, Nolan A Wages, and Ruitao Lin. “SFU: Surface-Free Utility-Based Design for Dose Optimization in Cancer Drug Combination Trials.” *Statistics in Biosciences* 16, no. 3 (2024): 854–881.
30. Peter F. Thall, Hoang Q. Nguyen, and Ralph G. Zinner. “Parametric Dose Standardization for Optimizing Two-Agent Combinations in a Phase I–II Trial with Ordinal Outcomes.” *Journal of the Royal Statistical Society, Series C (Applied Statistics)* 66, no. 1 (2017): 201–224.
31. Nolan A. Wages, and Mark R. Conaway. “Specifications of a Continual Reassessment Method Design for Phase I Trials of Combined Drugs.” *Pharmaceutical Statistics* 12, no. 4 (2013): 217–224.
32. Marie-Karelle Riviere, Frédéric Dubois, and Sarah Zohar. “Competing Designs for Drug Combination in Phase I Dose-Finding Clinical Trials.” *Statistics in Medicine* 34, no. 1 (2015): 1–12.
33. ICH 2024, August). “ICH M12 Guideline on Drug Interaction Studies. International Council for Harmonisation of Technical Requirements for Pharmaceuticals for Human Use (ICH).” ICH Step 4, endorsed by the ICH Assembly May 2024; FDA guidance availability announced 9 October 2024.
34. G. Bril, R. Dykstra, C. Pillers, and T. Robertson. “Algorithm AS 206: Isotonic Regression in Two Independent Variables.” *Journal of the Royal Statistical Society, Series C* 33, no. 3 (1984): 352–357.
35. Aditya Bardia, Shanu Modi, Mafalda Oliveira, Javier Cortes, Mario Campone, Brigette Ma, Luc Dirix, Amy Weise, Becker Hewes, Ivan Diaz-Padilla, et al. “Phase Ib dose-escalation/expansion trial of ribociclib in combination with everolimus and exemestane in postmenopausal women with HR+, HER2- advanced breast cancer.” *Clinical Cancer Research* 26, no. 24 (2020): 6417–6428.
36. Tim Robertson, F. T. Wright, and Richard L. Dykstra. 1988. *Order Restricted Statistical Inference*, New York: Wiley.
37. Ying-Kuen K Cheung. “Coherence in Phase I Clinical Trials.” *Methods and Applications of Statistics in Clinical Trials: Concepts, Principles, Trials, and Design* (2014): 230–239.
38. Ying Yuan, Sheng Liu, and Ronghui Lin. “Bayesian Optimal Interval Designs for Phase I Drug–Drug Combination Trials.” *Journal of the American Statistical Association* 113, no. 523 (2018): 1062–1073.
39. Anastasios Stathis, Anthony W. Tolcher, Judy S. Wang, Daniel J. Renouf, Lie-Chi Chen, Lauren H. Suttner, Tomoko Freshwater, Aimee L. Webber, Tapan Nayak, and Lillian L. Siu. “Results of an Open-Label Phase 1b Study of the ERK Inhibitor MK-8353 Plus the MEK Inhibitor Selumetinib in Patients with Advanced or Metastatic Solid Tumors.” *Investigational New Drugs* 41, no. 3 (2023): 380–390. ClinicalTrials.gov identifier NCT03745989.

# APPENDIX

## A List of Abbreviations

Table A1 lists all abbreviations used in the manuscript in full.

**Table A1.** Abbreviations used in the manuscript.

| Abbreviation | Full form |
|---|---|
| AL | Approximated likelihood |
| AUC | Area under the concentration–time curve |
| BOIN | Bayesian optimal interval |
| CDF | Cumulative distribution function |
| CL | Clearance |
| CRM | Continual reassessment method |
| CV | Coefficient of variation |
| DDI | Drug–drug interaction |
| DLT | Dose-limiting toxicity |
| ERK | Extracellular signal-regulated kinase |
| FDA | U.S. Food and Drug Administration |
| ICH | International Council for Harmonisation of Technical Requirements for Pharmaceuticals for Human Use |
| MEK | Mitogen-activated protein kinase kinase |
| MIDD | Model-informed drug development |
| MTD | Maximum tolerated dose |
| MTDC | Maximum tolerated dose combination |
| OBD | Optimal biological dose |
| OBDC | Optimal biological dose combination |
| PAVA | Pool-adjacent-violators algorithm |
| PBPK | Physiologically based pharmacokinetic |
| PK | Pharmacokinetic(s) |
| PMU | Posterior-mean utility |
| RDS | Rank-based desirability score |
| SAP | Statistical analysis plan |
| TITE | Time-to-event |

SUPPLEMENTARY MATERIAL

# PKComb-BOIN12: A Pharmacokinetically Guided Bayesian Design for Dose Optimisation in Oncology Drug-Combination Trials

Jieqi Tu, Ruitao Lin, and Ayon Mukherjee

## Overview

This Supplementary Material accompanies the main manuscript and provides: (S1) full dose-toxicity–efficacy–PK specifications for simulation scenarios S1–S19; (S2) full specifications and operating-characteristic breakdowns for the deliberately non-monotone assumption-violation scenarios S20–S21; (S3) the complete four-arm factorial decomposition (correct selection, overdose selection, early termination, admissibility, and coherence) underlying Table 2 of the main text; (S4) full per-scenario trial-duration reduction figures for the time-to-event (TITE) extension; (S5) the complete 19-scenario comparison of the proposed down-set PK-elimination rule against the alternative total-order ($j + k$) rule; (S6) the full early-termination breakdown underlying the target-transportability sensitivity analysis of Table 6; (S7) the complete $3 \times 3$ PK-sufficiency threshold ($\zeta_1$) by confidence-level ($C_P$) sensitivity grid; and (S8) the complete time-to-event robustness results under alternative event-time distributions and toxicity–efficacy dependence structures. All figures reported here were produced by the same audited simulation pipeline, available at https://github.com/jtucode/PKComb-BOIN12/, that generated every operating characteristic reported in the main text; every percentage is based on 5 000 simulated trials per design-scenario(-condition) combination, as stated in each table. Section, table, and equation numbers in this document are prefixed with "S" and are distinct from those of the main text; cross-references to the main text use its own (unprefixed) numbering.

## S1 Full Scenario Specifications: S1–S19

Table S1 (reproduced below scenario-by-scenario, as the full set of toxicity, efficacy, and PK matrices does not fit a single tabular block) gives the complete dose-toxicity–efficacy–PK specification for each of the nineteen base simulation scenarios used throughout Section 4 of the main text, exactly as used by the simulation pipeline (that is, after the minimal monotonicity correction described in Section 4.2 of the main text is applied to the two scenarios, S12 and S14, whose publicly released PK matrices contained a small number of internally non-monotone cells). $p_{j,k}$, $q_{j,k}$, and $r_{j,k}$ denote the true toxicity probability, efficacy (disease-control) probability, and mean PK exposure at the combination of Agent A dose $j$ and Agent B dose $k$ respectively; rows index Agent A doses (increasing downward) and columns index Agent B doses (increasing rightward), matching the convention of Section 2 of the main text. Target toxicity rate $p_T$, target efficacy rate $q_E$, and maximum per-scenario sample size $N$ are as stated in each scenario's heading.

**Table S1.** Full dose-toxicity–efficacy–PK specifications, Scenarios S1–S19

### Scenario S1 ($J = 4$, $K = 4$, $p_T = 0.35$, $q_E = 0.20$, $N = 51$)

Agent A doses: 80, 160, 240, 320. Agent B doses: 75, 150, 225, 300.

**Toxicity** $p_{j,k}$

| | | | | |
|---|---|---|---|---|
| A=80 | 0.07 | 0.09 | 0.12 | 0.13 |
| A=160 | 0.10 | 0.11 | 0.14 | 0.15 |
| A=240 | 0.12 | 0.14 | 0.19 | 0.20 |
| A=320 | 0.14 | 0.15 | 0.20 | 0.22 |

**Efficacy** $q_{j,k}$

| | | | | |
|---|---|---|---|---|
| A=80 | 0.13 | 0.20 | 0.20 | 0.30 |
| A=160 | 0.20 | 0.25 | 0.30 | 0.40 |
| A=240 | 0.25 | 0.30 | 0.40 | 0.60 |
| A=320 | 0.15 | 0.20 | 0.25 | 0.40 |

**PK** $r_{j,k}$

| | | | | |
|---|---|---|---|---|
| A=80 | 1000 | 1800 | 2200 | 3000 |
| A=160 | 1800 | 2500 | 3500 | 4200 |
| A=240 | 2000 | 3500 | 4500 | 6500 |
| A=320 | 4200 | 5000 | 6000 | 7000 |

### Scenario S2 ($J = 4$, $K = 4$, $p_T = 0.35$, $q_E = 0.20$, $N = 51$)

Agent A doses: 80, 160, 240, 320. Agent B doses: 75, 150, 225, 300.

**Toxicity** $p_{j,k}$

| | | | | |
|---|---|---|---|---|
| A=80 | 0.08 | 0.10 | 0.12 | 0.14 |
| A=160 | 0.10 | 0.12 | 0.15 | 0.18 |
| A=240 | 0.12 | 0.15 | 0.18 | 0.21 |
| A=320 | 0.15 | 0.18 | 0.20 | 0.30 |

**Efficacy** $q_{j,k}$

| | | | | |
|---|---|---|---|---|
| A=80 | 0.05 | 0.20 | 0.22 | 0.24 |
| A=160 | 0.10 | 0.25 | 0.28 | 0.30 |
| A=240 | 0.15 | 0.30 | 0.40 | 0.32 |
| A=320 | 0.20 | 0.35 | 0.60 | 0.40 |

**PK** $r_{j,k}$

| | | | | |
|---|---|---|---|---|
| A=80 | 1000 | 1800 | 2500 | 3200 |
| A=160 | 1800 | 2500 | 3500 | 3500 |
| A=240 | 2200 | 3500 | 5000 | 6000 |
| A=320 | 3500 | 4000 | 5000 | 6500 |

### Scenario S3 ($J = 4$, $K = 4$, $p_T = 0.35$, $q_E = 0.20$, $N = 51$)

Agent A doses: 80, 160, 240, 320. Agent B doses: 75, 150, 225, 300.

**Toxicity** $p_{j,k}$

| | | | | |
|---|---|---|---|---|
| A=80 | 0.18 | 0.20 | 0.24 | 0.30 |
| A=160 | 0.20 | 0.24 | 0.33 | 0.35 |
| A=240 | 0.25 | 0.28 | 0.32 | 0.40 |
| A=320 | 0.28 | 0.35 | 0.40 | 0.54 |

**Efficacy** $q_{j,k}$

| | | | | |
|---|---|---|---|---|
| A=80 | 0.20 | 0.25 | 0.50 | 0.40 |
| A=160 | 0.24 | 0.30 | 0.40 | 0.35 |
| A=240 | 0.28 | 0.32 | 0.35 | 0.30 |
| A=320 | 0.30 | 0.35 | 0.30 | 0.25 |

**PK** $r_{j,k}$

| | | | | |
|---|---|---|---|---|
| A=80 | 4000 | 4500 | 6500 | 7000 |
| A=160 | 4200 | 4800 | 6800 | 7200 |
| A=240 | 5000 | 5200 | 7000 | 7500 |
| A=320 | 5000 | 5500 | 7500 | 8000 |

### Scenario S4 ($J = 4$, $K = 4$, $p_T = 0.35$, $q_E = 0.20$, $N = 51$)

Agent A doses: 80, 160, 240, 320. Agent B doses: 75, 150, 225, 300.

**Toxicity** $p_{j,k}$

| | | | | |
|---|---|---|---|---|
| A=80 | 0.10 | 0.18 | 0.20 | 0.28 |
| A=160 | 0.20 | 0.22 | 0.35 | 0.48 |
| A=240 | 0.28 | 0.35 | 0.48 | 0.55 |
| A=320 | 0.32 | 0.45 | 0.56 | 0.60 |

**Efficacy** $q_{j,k}$

| | | | | |
|---|---|---|---|---|
| A=80 | 0.20 | 0.20 | 0.40 | 0.42 |
| A=160 | 0.20 | 0.24 | 0.55 | 0.58 |
| A=240 | 0.28 | 0.30 | 0.58 | 0.62 |
| A=320 | 0.25 | 0.32 | 0.35 | 0.62 |

**PK** $r_{j,k}$

| | | | | |
|---|---|---|---|---|
| A=80 | 2000 | 3000 | 4000 | 5000 |
| A=160 | 3000 | 4000 | 6000 | 7500 |
| A=240 | 4000 | 5500 | 7500 | 8000 |
| A=320 | 5000 | 6500 | 8500 | 9000 |

### Scenario S5 ($J = 4$, $K = 4$, $p_T = 0.35$, $q_E = 0.20$, $N = 51$)

Agent A doses: 80, 160, 240, 320. Agent B doses: 75, 150, 225, 300.

**Toxicity** $p_{j,k}$

| | | | | |
|---|---|---|---|---|
| A=80 | 0.12 | 0.16 | 0.18 | 0.20 |
| A=160 | 0.14 | 0.18 | 0.20 | 0.42 |
| A=240 | 0.18 | 0.20 | 0.42 | 0.55 |
| A=320 | 0.20 | 0.40 | 0.52 | 0.60 |

**Efficacy** $q_{j,k}$

| | | | | |
|---|---|---|---|---|
| A=80 | 0.20 | 0.25 | 0.30 | 0.50 |
| A=160 | 0.20 | 0.25 | 0.30 | 0.60 |
| A=240 | 0.30 | 0.50 | 0.60 | 0.66 |
| A=320 | 0.40 | 0.60 | 0.65 | 0.70 |

**PK** $r_{j,k}$

| | | | | |
|---|---|---|---|---|
| A=80 | 2000 | 3000 | 4000 | 5500 |
| A=160 | 2000 | 3000 | 4000 | 5500 |
| A=240 | 4000 | 5500 | 7000 | 8000 |
| A=320 | 5000 | 7000 | 8500 | 9000 |

### Scenario S6 ($J = 4$, $K = 4$, $p_T = 0.35$, $q_E = 0.20$, $N = 51$)

Agent A doses: 80, 160, 240, 320. Agent B doses: 75, 150, 225, 300.

**Toxicity** $p_{j,k}$

| | | | | |
|---|---|---|---|---|
| A=80 | 0.05 | 0.10 | 0.12 | 0.15 |
| A=160 | 0.07 | 0.11 | 0.14 | 0.17 |
| A=240 | 0.10 | 0.12 | 0.15 | 0.19 |
| A=320 | 0.12 | 0.15 | 0.16 | 0.23 |

**Efficacy** $q_{j,k}$

| | | | | |
|---|---|---|---|---|
| A=80 | 0.15 | 0.16 | 0.20 | 0.24 |
| A=160 | 0.18 | 0.22 | 0.28 | 0.40 |
| A=240 | 0.20 | 0.26 | 0.32 | 0.45 |
| A=320 | 0.23 | 0.28 | 0.35 | 0.60 |

**PK** $r_{j,k}$

| | | | | |
|---|---|---|---|---|
| A=80 | 1000 | 2000 | 2500 | 3000 |
| A=160 | 2500 | 3000 | 3500 | 4000 |
| A=240 | 3000 | 3500 | 4000 | 4500 |
| A=320 | 3500 | 4000 | 4000 | 6500 |

**Scenario S7 ($J = 4, K = 4, p_T = 0.35, q_E = 0.20, N = 51$)**

Agent A doses: 80, 160, 240, 320. Agent B doses: 75, 150, 225, 300.

| **Toxicity** $p_{j,k}$ | | | | |
|---|---|---|---|---|
| A=80 | 0.08 | 0.14 | 0.18 | 0.35 |
| A=160 | 0.10 | 0.16 | 0.21 | 0.42 |
| A=240 | 0.12 | 0.18 | 0.24 | 0.45 |
| A=320 | 0.16 | 0.24 | 0.33 | 0.60 |

| **Efficacy** $q_{j,k}$ | | | | |
|---|---|---|---|---|
| A=80 | 0.10 | 0.15 | 0.25 | 0.55 |
| A=160 | 0.12 | 0.20 | 0.30 | 0.55 |
| A=240 | 0.16 | 0.26 | 0.55 | 0.55 |
| A=320 | 0.20 | 0.30 | 0.55 | 0.55 |

| **PK** $r_{j,k}$ | | | | |
|---|---|---|---|---|
| A=80 | 1500 | 2500 | 3500 | 5000 |
| A=160 | 2500 | 3500 | 4500 | 5500 |
| A=240 | 3500 | 4500 | 5800 | 6500 |
| A=320 | 4500 | 5500 | 6500 | 7500 |

**Scenario S8 ($J = 4, K = 4, p_T = 0.35, q_E = 0.20, N = 51$)**

Agent A doses: 80, 160, 240, 320. Agent B doses: 75, 150, 225, 300.

| **Toxicity** $p_{j,k}$ | | | | |
|---|---|---|---|---|
| A=80 | 0.01 | 0.05 | 0.12 | 0.15 |
| A=160 | 0.03 | 0.08 | 0.16 | 0.20 |
| A=240 | 0.06 | 0.12 | 0.20 | 0.26 |
| A=320 | 0.10 | 0.15 | 0.23 | 0.33 |

| **Efficacy** $q_{j,k}$ | | | | |
|---|---|---|---|---|
| A=80 | 0.20 | 0.10 | 0.15 | 0.15 |
| A=160 | 0.30 | 0.25 | 0.25 | 0.20 |
| A=240 | 0.55 | 0.35 | 0.30 | 0.25 |
| A=320 | 0.35 | 0.25 | 0.20 | 0.15 |

| **PK** $r_{j,k}$ | | | | |
|---|---|---|---|---|
| A=80 | 2000 | 2100 | 2200 | 2250 |
| A=160 | 4000 | 4100 | 4200 | 4300 |
| A=240 | 7000 | 7100 | 7120 | 7300 |
| A=320 | 7100 | 7200 | 7300 | 7400 |

**Scenario S9 ($J = 4, K = 4, p_T = 0.35, q_E = 0.20, N = 51$)**

Agent A doses: 80, 160, 240, 320. Agent B doses: 75, 150, 225, 300.

| **Toxicity** $p_{j,k}$ | | | | |
|---|---|---|---|---|
| A=80 | 0.50 | 0.56 | 0.65 | 0.68 |
| A=160 | 0.55 | 0.62 | 0.69 | 0.71 |
| A=240 | 0.60 | 0.67 | 0.75 | 0.79 |
| A=320 | 0.62 | 0.72 | 0.80 | 0.85 |

| **Efficacy** $q_{j,k}$ | | | | |
|---|---|---|---|---|
| A=80 | 0.52 | 0.62 | 0.70 | 0.70 |
| A=160 | 0.55 | 0.66 | 0.74 | 0.79 |
| A=240 | 0.58 | 0.70 | 0.78 | 0.82 |
| A=320 | 0.65 | 0.74 | 0.80 | 0.88 |

| **PK** $r_{j,k}$ | | | | |
|---|---|---|---|---|
| A=80 | 2000 | 2050 | 2080 | 2100 |
| A=160 | 3500 | 3580 | 3580 | 3600 |
| A=240 | 4100 | 4170 | 4790 | 4790 |
| A=320 | 6000 | 6350 | 6550 | 6800 |

**Scenario S10 ($J = 4, K = 4, p_T = 0.30, q_E = 0.20, N = 45$)**

Agent A doses: 80, 160, 240, 320. Agent B doses: 75, 150, 225, 300.

| **Toxicity** $p_{j,k}$ | | | | |
|---|---|---|---|---|
| A=80 | 0.02 | 0.04 | 0.07 | 0.12 |
| A=160 | 0.04 | 0.08 | 0.13 | 0.18 |
| A=240 | 0.09 | 0.15 | 0.18 | 0.25 |
| A=320 | 0.14 | 0.14 | 0.18 | 0.25 |

| **Efficacy** $q_{j,k}$ | | | | |
|---|---|---|---|---|
| A=80 | 0.08 | 0.10 | 0.29 | 0.42 |
| A=160 | 0.10 | 0.23 | 0.42 | 0.60 |
| A=240 | 0.14 | 0.14 | 0.24 | 0.43 |
| A=320 | 0.10 | 0.10 | 0.10 | 0.18 |

| **PK** $r_{j,k}$ | | | | |
|---|---|---|---|---|
| A=80 | 800 | 1000 | 3000 | 4000 |
| A=160 | 1200 | 4000 | 4200 | 6800 |
| A=240 | 3100 | 4000 | 4200 | 6830 |
| A=320 | 3200 | 4000 | 4400 | 6850 |

**Scenario S11 ($J = 4, K = 4, p_T = 0.30, q_E = 0.20, N = 45$)**

Agent A doses: 80, 160, 240, 320. Agent B doses: 75, 150, 225, 300.

| **Toxicity** $p_{j,k}$ | | | | |
|---|---|---|---|---|
| A=80 | 0.01 | 0.07 | 0.12 | 0.18 |
| A=160 | 0.05 | 0.12 | 0.15 | 0.20 |
| A=240 | 0.10 | 0.19 | 0.20 | 0.21 |
| A=320 | 0.15 | 0.18 | 0.21 | 0.25 |

| **Efficacy** $q_{j,k}$ | | | | |
|---|---|---|---|---|
| A=80 | 0.05 | 0.22 | 0.15 | 0.08 |
| A=160 | 0.12 | 0.29 | 0.15 | 0.10 |
| A=240 | 0.19 | 0.44 | 0.20 | 0.18 |
| A=320 | 0.42 | 0.60 | 0.38 | 0.32 |

| **PK** $r_{j,k}$ | | | | |
|---|---|---|---|---|
| A=80 | 500 | 2500 | 2500 | 3000 |
| A=160 | 1000 | 2500 | 3500 | 4000 |
| A=240 | 2000 | 3000 | 4500 | 5000 |
| A=320 | 4000 | 7500 | 7600 | 7800 |

**Scenario S12 ($J = 3, K = 4, p_T = 0.30, q_E = 0.20, N = 45$)**

Agent A doses: 80, 160, 240. Agent B doses: 75, 150, 225, 300.

| **Toxicity** $p_{j,k}$ | | | | |
|---|---|---|---|---|
| A=80 | 0.05 | 0.10 | 0.18 | 0.25 |
| A=160 | 0.05 | 0.10 | 0.23 | 0.42 |
| A=240 | 0.15 | 0.23 | 0.42 | 0.43 |

| **Efficacy** $q_{j,k}$ | | | | |
|---|---|---|---|---|
| A=80 | 0.30 | 0.40 | 0.60 | 0.37 |
| A=160 | 0.20 | 0.28 | 0.37 | 0.26 |
| A=240 | 0.10 | 0.14 | 0.24 | 0.18 |

| **PK** $r_{j,k}$ | | | | |
|---|---|---|---|---|
| A=80 | 3000 | 4500 | 5500 | 6000 |
| A=160 | 3200 | 4510 | 5520 | 6000 |
| A=240 | 3300 | 4550 | 5530 | 6000 |

**Scenario S13 ($J = 4, K = 4, p_T = 0.30, q_E = 0.20, N = 45$)**

Agent A doses: 80, 160, 240, 320. Agent B doses: 75, 150, 225, 300.

| **Toxicity** $p_{j,k}$ | | | | |
|---|---|---|---|---|
| A=80 | 0.05 | 0.18 | 0.26 | 0.38 |
| A=160 | 0.15 | 0.26 | 0.40 | 0.49 |
| A=240 | 0.25 | 0.42 | 0.46 | 0.51 |
| A=320 | 0.39 | 0.45 | 0.50 | 0.49 |

| **Efficacy** $q_{j,k}$ | | | | |
|---|---|---|---|---|
| A=80 | 0.60 | 0.37 | 0.30 | 0.24 |
| A=160 | 0.30 | 0.22 | 0.26 | 0.13 |
| A=240 | 0.24 | 0.18 | 0.14 | 0.10 |
| A=320 | 0.15 | 0.10 | 0.08 | 0.05 |

| **PK** $r_{j,k}$ | | | | |
|---|---|---|---|---|
| A=80 | 6000 | 6200 | 6400 | 6600 |
| A=160 | 6400 | 6600 | 6800 | 6800 |
| A=240 | 6400 | 6600 | 6800 | 6850 |
| A=320 | 6400 | 6600 | 6900 | 7000 |

**Scenario S14 ($J = 4, K = 4, p_T = 0.30, q_E = 0.20, N = 45$)**

Agent A doses: 80, 160, 240, 320. Agent B doses: 75, 150, 225, 300.

| **Toxicity** $p_{j,k}$ | | | | |
|---|---|---|---|---|
| A=80 | 0.09 | 0.17 | 0.22 | 0.24 |
| A=160 | 0.17 | 0.22 | 0.24 | 0.25 |
| A=240 | 0.22 | 0.24 | 0.25 | 0.25 |
| A=320 | 0.24 | 0.25 | 0.25 | 0.25 |

| **Efficacy** $q_{j,k}$ | | | | |
|---|---|---|---|---|
| A=80 | 0.00 | 0.11 | 0.23 | 0.11 |
| A=160 | 0.01 | 0.14 | 0.26 | 0.13 |
| A=240 | 0.01 | 0.23 | 0.39 | 0.22 |
| A=320 | 0.25 | 0.06 | 0.62 | 0.42 |

| **PK** $r_{j,k}$ | | | | |
|---|---|---|---|---|
| A=80 | 500 | 1500 | 3000 | 3500 |
| A=160 | 500 | 3000 | 3500 | 3510 |
| A=240 | 600 | 4000 | 4200 | 8000 |
| A=320 | 800 | 4100 | 4200 | 8100 |

**Scenario S15 ($J = 4, K = 4, p_T = 0.30, q_E = 0.20, N = 45$)**

Agent A doses: 80, 160, 240, 320. Agent B doses: 75, 150, 225, 300.

| **Toxicity** $p_{j,k}$ | | | | |
|---|---|---|---|---|
| A=80 | 0.01 | 0.01 | 0.07 | 0.24 |
| A=160 | 0.01 | 0.07 | 0.23 | 0.39 |
| A=240 | 0.06 | 0.22 | 0.38 | 0.44 |
| A=320 | 0.21 | 0.38 | 0.44 | 0.45 |

| **Efficacy** $q_{j,k}$ | | | | |
|---|---|---|---|---|
| A=80 | 0.06 | 0.45 | 0.63 | 0.45 |
| A=160 | 0.03 | 0.32 | 0.50 | 0.32 |
| A=240 | 0.00 | 0.08 | 0.18 | 0.08 |
| A=320 | 0.00 | 0.00 | 0.01 | 0.00 |

| **PK** $r_{j,k}$ | | | | |
|---|---|---|---|---|
| A=80 | 200 | 4000 | 6000 | 6000 |
| A=160 | 200 | 4000 | 6000 | 6000 |
| A=240 | 200 | 4000 | 6000 | 6000 |
| A=320 | 200 | 4000 | 6000 | 6000 |

**Scenario S16 ($J = 2, K = 3, p_T = 0.33, q_E = 0.20, N = 42$)**

Agent A doses: 100, 200. Agent B doses: 50, 100, 200.

| **Toxicity** $p_{j,k}$ | | | |
|---|---|---|---|
| A=100 | 0.05 | 0.15 | 0.20 |
| A=200 | 0.10 | 0.15 | 0.45 |

| **Efficacy** $q_{j,k}$ | | | |
|---|---|---|---|
| A=100 | 0.10 | 0.30 | 0.50 |
| A=200 | 0.20 | 0.40 | 0.60 |

| **PK** $r_{j,k}$ | | | |
|---|---|---|---|
| A=100 | 1000 | 4500 | 5500 |
| A=200 | 3500 | 5000 | 6000 |

**Scenario S17 ($J = 2, K = 3, p_T = 0.33, q_E = 0.20, N = 42$)**

Agent A doses: 100, 200. Agent B doses: 50, 100, 200.

| **Toxicity** $p_{j,k}$ | | | |
|---|---|---|---|
| A=100 | 0.05 | 0.15 | 0.40 |
| A=200 | 0.10 | 0.20 | 0.50 |

| **Efficacy** $q_{j,k}$ | | | |
|---|---|---|---|
| A=100 | 0.10 | 0.30 | 0.50 |
| A=200 | 0.20 | 0.40 | 0.55 |

| **PK** $r_{j,k}$ | | | |
|---|---|---|---|
| A=100 | 2000 | 3000 | 4500 |
| A=200 | 3000 | 5500 | 7000 |

**Scenario S18 ($J = 2, K = 3, p_T = 0.33, q_E = 0.20, N = 42$)**

Agent A doses: 100, 200. Agent B doses: 50, 100, 200.

| **Toxicity** $p_{j,k}$ | | | |
|---|---|---|---|
| A=100 | 0.05 | 0.10 | 0.15 |
| A=200 | 0.10 | 0.15 | 0.20 |

| **Efficacy** $q_{j,k}$ | | | |
|---|---|---|---|
| A=100 | 0.10 | 0.20 | 0.40 |
| A=200 | 0.20 | 0.30 | 0.50 |

| **PK** $r_{j,k}$ | | | |
|---|---|---|---|
| A=100 | 2000 | 4000 | 5000 |
| A=200 | 3000 | 4500 | 7000 |

**Scenario S19 ($J = 2, K = 3, p_T = 0.33, q_E = 0.20, N = 42$)**

Agent A doses: 100, 200. Agent B doses: 50, 100, 200.

| **Toxicity** $p_{j,k}$ | | | |
|---|---|---|---|
| A=100 | 0.05 | 0.05 | 0.05 |
| A=200 | 0.05 | 0.05 | 0.05 |

| **Efficacy** $q_{j,k}$ | | | |
|---|---|---|---|
| A=100 | 0.10 | 0.20 | 0.40 |
| A=200 | 0.20 | 0.30 | 0.50 |

| **PK** $r_{j,k}$ | | | |
|---|---|---|---|
| A=100 | 2000 | 3000 | 4000 |
| A=200 | 3000 | 4500 | 6000 |

## S2 Non-Monotone PK Assumption-Violation Scenarios: S20–S21

S20 and S21 (Section 4.6.2 of the main text) are built from the *unpatched*, raw PK matrices of base scenarios S2 and S12 respectively, deliberately retaining the small number of internally non-monotone PK cells that were corrected (via the smallest possible isotonic raise) for S1–S19. This directly evaluates robustness to a local violation of Assumption 1 (monotonicity of PK exposure in each agent's dose). The non-monotone cell(s) are italicised in each PK matrix below. Table S2 gives the complete operating-characteristic breakdown (5 000 simulated trials per design-scenario pair) underlying Table 4 of the main text, including early-termination percentage, admissibility, coherence, and the mean number of replications retaining a true OBDC or selecting an overdose combination in the final admissible set, neither of which is shown in the main-text table for reasons of space.

**Table S2a.** Full dose-toxicity–efficacy–PK specifications, Scenarios S20–S21

**Scenario S20 (built from S2's raw, unpatched grid,** $J = 4$, $K = 4$, $p_T = 0.35$, $q_E = 0.20$, $N = 51$**)**

Agent A doses: 80, 160, 240, 320. Agent B doses: 75, 150, 225, 300.

| **Toxicity** $p_{j,k}$ | | | | |
|---|---|---|---|---|
| A=80 | 0.08 | 0.10 | 0.12 | 0.14 |
| A=160 | 0.10 | 0.12 | 0.15 | 0.18 |
| A=240 | 0.12 | 0.15 | 0.18 | 0.21 |
| A=320 | 0.15 | 0.18 | 0.20 | 0.30 |

| **Efficacy** $q_{j,k}$ | | | | |
|---|---|---|---|---|
| A=80 | 0.05 | 0.20 | 0.22 | 0.24 |
| A=160 | 0.10 | 0.25 | 0.28 | 0.30 |
| A=240 | 0.15 | 0.30 | 0.40 | 0.32 |
| A=320 | 0.20 | 0.35 | 0.60 | 0.40 |

| **PK** $r_{j,k}$ (*deliberately non-monotone cells in italics*) | | | | |
|---|---|---|---|---|
| A=80 | 1000 | 1800 | 2500 | 3200 |
| A=160 | 1800 | 2500 | 3500 | 3500 |
| A=240 | 2200 | 3500 | 5000 | 6000 |
| A=320 | 3500 | 4000 | *4500* | 6500 |

**Scenario S21 (built from S12's raw, unpatched grid,** $J = 3$, $K = 4$, $p_T = 0.30$, $q_E = 0.20$, $N = 45$**)**

Agent A doses: 80, 160, 240. Agent B doses: 75, 150, 225, 300.

| **Toxicity** $p_{j,k}$ | | | | |
|---|---|---|---|---|
| A=80 | 0.05 | 0.10 | 0.18 | 0.25 |
| A=160 | 0.05 | 0.10 | 0.23 | 0.42 |
| A=240 | 0.15 | 0.23 | 0.42 | 0.43 |

| **Efficacy** $q_{j,k}$ | | | | |
|---|---|---|---|---|
| A=80 | 0.30 | 0.40 | 0.60 | 0.37 |
| A=160 | 0.20 | 0.28 | 0.37 | 0.26 |
| A=240 | 0.10 | 0.14 | 0.24 | 0.18 |

| **PK** $r_{j,k}$ (*deliberately non-monotone cells in italics*) | | | | |
|---|---|---|---|---|
| A=80 | 3000 | 4500 | 5500 | 6000 |
| A=160 | 3200 | 4510 | 5520 | *5560* |
| A=240 | 3300 | 4550 | 5530 | 5570 |

**Table S2b.** Complete operating-characteristic breakdown, Scenarios S20–S21 (5 000 simulated trials per design-scenario pair)

| Scenario | Design | Correct % | Overdose % | Early % | Admiss. % | Coher. % | Mean $n$ true OBDC | Mean $n$ overdose |
|---|---|---|---|---|---|---|---|---|
| S20 | Comb-BOIN12 | 14.3 | 0.0 | 2.1 | 72.4 | 81.6 | 4.29 | 0.00 |
| S20 | PKComb-BOIN12 | 13.0 | 0.0 | 34.6 | 23.9 | 90.9 | 3.68 | 0.00 |
| S21 | Comb-BOIN12 | 55.0 | 1.8 | 1.0 | 84.6 | 85.8 | 10.79 | 2.06 |
| S21 | PKComb-BOIN12 | 63.8 | 3.3 | 8.8 | 80.3 | 88.7 | 11.98 | 3.46 |

## S3 Complete Factorial Decomposition

Table S3 reports the complete four-arm factorial decomposition (Section 4.4 of the main text) for all six structurally diverse scenarios (S1, S5, S8, S9, S14, S17), including admissibility and coherence percentages alongside the correct-selection, overdose-selection, and early-termination percentages shown in Table 2 of the main text (5 000 simulated trials per design-scenario pair). D0 = Comb-BOIN12; D1 = Comb-BOIN12 with the final admissible set additionally PK-restricted (estimand redefinition only); D2 = D1 with the PK-based posterior-mean-utility final estimator (estimator change); D3 = PKComb-BOIN12 (adds real-time PK-gated escalation and PK elimination). Admissibility is the percentage of simulated trials whose final admissible set retains a truly admissible combination when one exists; coherence is the percentage of dose-assignment transitions consistent with the observed toxicity/efficacy/PK data at each step (Cheung, 2014, *Coherence in Phase I Clinical Trials*). S9 is a fully toxic scenario with no true admissible combination, so admissibility is undefined for S9 (shown as "–").

**Table S3.** Complete four-arm factorial decomposition, S1, S5, S8, S9, S14, S17 (5 000 simulated trials per design-scenario pair)

| Scenario | Design | Correct % | Overdose % | Early % | Admiss. % | Coher. % |
|---|---|---|---|---|---|---|
| S1 | D0 | 16.2 | 0.0 | 1.3 | 26.9 | 80.0 |
| S1 | D1 | 21.0 | 0.0 | 1.3 | 26.4 | 80.0 |
| S1 | D2 | 20.2 | 0.0 | 1.3 | 26.4 | 80.0 |
| S1 | D3 | 55.9 | 0.0 | 22.0 | 68.8 | 89.1 |
| S5 | D0 | 31.1 | 24.1 | 4.7 | 84.3 | 89.9 |
| S5 | D1 | 43.1 | 29.7 | 4.7 | 74.9 | 89.9 |
| S5 | D2 | 44.4 | 29.8 | 4.7 | 74.9 | 89.9 |
| S5 | D3 | 41.7 | 29.5 | 29.3 | 70.4 | 93.9 |
| S8 | D0 | 50.1 | 0.0 | 0.1 | 99.6 | 88.7 |
| S8 | D1 | 59.6 | 0.0 | 0.1 | 98.3 | 88.7 |
| S8 | D2 | 62.6 | 0.0 | 0.1 | 98.3 | 88.7 |
| S8 | D3 | 64.3 | 0.0 | 4.6 | 98.7 | 89.6 |
| S9[†] | D0 | 0.0 | 36.2 | 95.9 | – | 93.4 |
| S9 | D1 | 0.0 | 0.1 | 95.9 | – | 93.4 |
| S9 | D2 | 0.0 | 0.1 | 95.9 | – | 93.4 |
| S9 | D3 | 0.0 | 0.3 | 100.0 | – | 100.0 |
| S14 | D0 | 1.9 | 0.0 | 3.8 | 5.1 | 81.9 |
| S14 | D1 | 3.4 | 0.0 | 3.8 | 5.1 | 81.9 |
| S14 | D2 | 3.9 | 0.0 | 3.8 | 5.1 | 81.9 |
| S14 | D3 | 20.0 | 0.0 | 57.0 | 25.0 | 94.1 |
| S17 | D0 | 30.5 | 28.4 | 1.0 | 92.8 | 89.5 |
| S17 | D1 | 72.0 | 13.8 | 1.0 | 83.1 | 89.5 |
| S17 | D2 | 72.6 | 13.2 | 1.0 | 83.1 | 89.5 |
| S17 | D3 | 71.0 | 13.8 | 56.6 | 81.6 | 90.9 |

[†]S9 is a fully toxic scenario with no true OBDC; admissibility is undefined (shown as –).

## S4 Time-to-Event Extension: Full Duration-Reduction Results

Table S4 gives the full per-scenario mean total-enrolment figures (used as a proxy for trial duration, Section 4.5 of the main text) for Comb-BOIN12 and PKComb-BOIN12 and their respective TITE variants, together with the percentage reduction achieved by adopting the TITE extension, across S1–S19 excluding S9 (5 000 simulated trials per design-scenario pair). "Mean $n$" is the mean number of patients enrolled per simulated trial.

**Table S4.** Full per-scenario trial-duration (mean enrolment) reduction under the TITE extension (5 000 simulated trials per design-scenario pair)

| Scenario | Comb-BOIN12 | | | PKComb-BOIN12 | | |
|---|---|---|---|---|---|---|
| | Mean $n$ | TITE mean $n$ | Reduction % | Mean $n$ | TITE mean $n$ | Reduction % |
| S1 | 58.75 | 29.75 | 49.4 | 53.82 | 26.27 | 51.2 |
| S2 | 58.39 | 29.29 | 49.8 | 52.51 | 25.41 | 51.6 |
| S3 | 53.95 | 26.57 | 50.8 | 49.17 | 24.48 | 50.2 |
| S4 | 57.81 | 28.45 | 50.8 | 43.51 | 22.62 | 48.0 |
| S5 | 56.97 | 28.35 | 50.2 | 50.38 | 25.09 | 50.2 |
| S6 | 59.11 | 30.01 | 49.2 | 53.11 | 25.38 | 52.2 |
| S7 | 58.45 | 29.06 | 50.3 | 54.03 | 26.31 | 51.3 |
| S8 | 59.46 | 29.36 | 50.6 | 58.85 | 28.25 | 52.0 |
| S10 | 52.42 | 26.49 | 49.5 | 51.09 | 24.39 | 52.3 |
| S11 | 52.47 | 25.60 | 51.2 | 48.33 | 21.66 | 55.2 |
| S12 | 52.18 | 24.02 | 54.0 | 50.64 | 21.41 | 57.7 |
| S13 | 52.05 | 22.48 | 56.8 | 52.04 | 22.45 | 56.9 |
| S14 | 51.30 | 20.70 | 59.6 | 42.01 | 15.23 | 63.8 |
| S15 | 52.48 | 25.08 | 52.2 | 51.98 | 23.69 | 54.4 |
| S16 | 48.71 | 24.22 | 50.3 | 46.24 | 22.26 | 51.9 |
| S17 | 48.63 | 23.99 | 50.7 | 38.88 | 19.32 | 50.3 |
| S18 | 48.69 | 24.38 | 49.9 | 46.92 | 22.56 | 51.9 |
| S19 | 48.67 | 24.82 | 49.0 | 48.28 | 23.15 | 52.0 |

Mean reduction: 51.3% (Comb-BOIN12), 52.9% (PKComb-BOIN12); range 48.0%–63.8% overall.

## S5 PK-Elimination Rule Comparison: Full 19-Scenario Results

Table S5 gives the complete per-scenario comparison of PKComb-BOIN12 under the proposed down-set PK-elimination rule against an otherwise identical design using the alternative total-order ($j + k$) rule discussed in Example 1 and Figure 1 of the main text (Section 4.6.1 of the main text), across all nineteen scenarios S1–S19 (5 000 replications per scenario-rule pair, independently simulated from the headline results of Table 1 of the main text using an independent pseudo-random stream; down-set-rule figures for a given scenario can therefore differ from Table 1 by up to a few tenths of a percentage point due to Monte Carlo noise, rather than any difference in the simulated design). ΔCorrect and ΔOverdose are the down-set rule's correct-selection and overdose-selection percentages minus the total-order rule's, computed from unrounded per-scenario percentages and then rounded to one decimal place for display, so a negative Δ Overdose indicates the down-set rule is safer.

**Table S5.** Down-set versus total-order ($j + k$) PK-elimination rule, full 19-scenario comparison (5 000 replications per scenario-rule pair)

| Scenario | Down-set rule | | Total-order ($j + k$) rule | | Difference (down-set − total-order) | |
|---|---|---|---|---|---|---|
| | Correct % | Overdose % | Correct % | Overdose % | ΔCorrect pp | ΔOverdose pp |
| S1 | 55.9 | 0.0 | 63.2 | 0.0 | −7.3 | 0.0 |
| S2 | 38.7 | 0.0 | 43.6 | 0.0 | −4.9 | 0.0 |
| S3 | 31.8 | 0.6 | 29.4 | 0.5 | +2.4 | 0.0 |
| S4 | 26.0 | 11.3 | 27.9 | 12.5 | −1.9 | −1.3 |
| S5 | 41.7 | 29.6 | 42.2 | 32.9 | −0.5 | −3.3 |
| S6 | 64.2 | 0.0 | 73.4 | 0.0 | −9.3 | 0.0 |
| S7 | 35.9 | 9.9 | 39.0 | 11.4 | −3.1 | −1.5 |
| S8 | 64.3 | 0.0 | 62.5 | 0.0 | +1.8 | 0.0 |
| S9† | 0.0 | 0.3 | 0.0 | 0.3 | 0.0 | 0.0 |
| S10 | 70.9 | 0.0 | 72.3 | 0.0 | −1.4 | 0.0 |
| S11 | 72.1 | 0.0 | 76.7 | 0.0 | −4.6 | 0.0 |
| S12 | 63.3 | 3.8 | 56.7 | 5.5 | +6.6 | −1.7 |
| S13 | 77.4 | 1.5 | 77.4 | 1.5 | 0.0 | 0.0 |
| S14 | 20.0 | 0.0 | 26.3 | 0.0 | −6.3 | 0.0 |
| S15 | 68.4 | 1.7 | 55.2 | 5.2 | +13.2 | −3.5 |
| S16 | 51.7 | 13.3 | 51.0 | 14.0 | +0.7 | −0.8 |
| S17 | 71.0 | 13.8 | 70.8 | 14.7 | +0.2 | −0.9 |
| S18 | 64.5 | 0.0 | 67.9 | 0.0 | −3.4 | 0.0 |
| S19 | 96.2 | 0.0 | 97.1 | 0.0 | −0.9 | 0.0 |

†S9 has no true OBDC and is excluded from the correct-selection summary statistics reported in the main text, but shown here for completeness.

## S6 Target-Transportability Sensitivity: Early-Termination Breakdown

Table S6 reproduces the correct-selection percentages of Table 6 of the main text alongside early-termination percentages, for the naive fixed-target design ($\tau_{\text{used}} = 1$) and Algorithm 1's adaptive calibration, under true transportability factors $\tau_{\text{true}} \in \{0.7, 1.0, 1.3\}$, in S1 and S7 (5 000 replications per scenario-value-mode triple; Section 4.6.4 of the main text).

**Table S6.** Target-transportability sensitivity: correct-selection and early-termination percentages, naive versus adaptive calibration

| $\tau_{\text{true}}$ | Mode | Scenario 1 | | Scenario 7 | |
|---|---|---|---|---|---|
| | | Correct % | Early % | Correct % | Early % |
| 0.7 | Naive | 8.8 | 42.3 | 0.5 | 79.2 |
| 0.7 | Adaptive | 59.4 | 27.9 | 37.5 | 31.5 |
| 1.0 | Naive | 55.9 | 22.0 | 35.9 | 26.2 |
| 1.0 | Adaptive | 56.5 | 21.5 | 36.2 | 26.0 |
| 1.3 | Naive | 24.7 | 13.2 | 25.3 | 13.1 |
| 1.3 | Adaptive | 49.2 | 19.9 | 31.2 | 26.0 |

## S7 PK-Sufficiency Threshold by Confidence-Level Sensitivity Grid

Table S7 gives the complete $3 \times 3$ sensitivity grid varying the PK-sufficiency threshold $\zeta_1 \in \{4200, 4800, 5400\}$ (equivalently $r_P \in \{5250, 6000, 6750\}$ at fixed $\zeta_1/r_P = 0.8$) against the PK-elimination confidence level $C_P \in \{0.90, 0.95, 0.99\}$, in Scenario 1 (5 000 replications per cell; Section 4.6.4 of the main text).

**Table S7.** $\zeta_1 \times C_P$ sensitivity grid, Scenario 1 (5 000 replications per cell)

| $\zeta_1$ | $C_P$ | Correct % | Overdose % | Early % | Admiss. % |
|---|---|---|---|---|---|
| 4200 | 0.90 | 38.1 | 0.0 | 18.3 | 54.6 |
| 4200 | 0.95 | 38.4 | 0.0 | 16.7 | 54.7 |
| 4200 | 0.99 | 38.3 | 0.0 | 13.7 | 54.5 |
| 4800 | 0.90 | 55.1 | 0.0 | 24.1 | 67.9 |
| 4800 | 0.95 | 55.9 | 0.0 | 22.0 | 68.8 |
| 4800 | 0.99 | 57.4 | 0.0 | 18.5 | 70.4 |
| 5400 | 0.90 | 57.2 | 0.0 | 31.6 | 66.2 |
| 5400 | 0.95 | 57.8 | 0.0 | 29.2 | 67.1 |
| 5400 | 0.99 | 60.2 | 0.0 | 24.3 | 69.2 |

## S8 TITE Robustness: Full Results by Event-Time Distribution and Dependence

Table S8 gives the complete results underlying the TITE robustness analysis of Section 4.6.5 of the main text: correct-selection, overdose-selection, early-termination, and admissibility percentages for TITE-PKComb-BOIN12 under event times drawn either uniformly or from a Weibull distribution with increasing hazard (shape parameter 1.8, matched in mean to the uniform case), crossed with toxicity–efficacy event-time dependence induced via a Gaussian copula with correlation $\rho \in \{0, 0.5\}$, in S1, S5, and S7 (5 000 replications per scenario-distribution-correlation combination).

**Table S8.** TITE robustness to event-time distribution and toxicity–efficacy dependence, full results (5 000 replications per scenario-distribution-correlation combination)

| Scenario | Distribution | $\rho$ | Correct % | Overdose % | Early % | Admiss. % |
|---|---|---|---|---|---|---|
| S1 | Uniform | 0.0 | 59.4 | 0.0 | 16.0 | 73.1 |
| S1 | Uniform | 0.5 | 60.6 | 0.0 | 19.2 | 72.8 |
| S1 | Weibull | 0.0 | 59.3 | 0.0 | 15.6 | 73.7 |
| S1 | Weibull | 0.5 | 61.3 | 0.0 | 18.2 | 74.2 |
| S5 | Uniform | 0.0 | 41.7 | 32.2 | 21.0 | 72.0 |
| S5 | Uniform | 0.5 | 43.4 | 29.8 | 26.4 | 72.9 |
| S5 | Weibull | 0.0 | 42.4 | 32.4 | 20.9 | 73.0 |
| S5 | Weibull | 0.5 | 44.4 | 29.9 | 24.7 | 74.2 |
| S7 | Uniform | 0.0 | 37.9 | 11.6 | 18.2 | 67.2 |
| S7 | Uniform | 0.5 | 39.6 | 9.6 | 21.9 | 67.7 |
| S7 | Weibull | 0.0 | 38.7 | 12.0 | 16.9 | 68.4 |
| S7 | Weibull | 0.5 | 39.7 | 11.1 | 21.2 | 68.4 |

## Software and Reproducibility

All simulation code used to produce the results in Tables S1–S8 (and in Tables 1–7 of the main text) is available at https://github.com/jtucode/PKComb-BOIN12/, together with a README describing how to reproduce each table from the scenario definitions (`scenarios.py`), design configurations (`design.py`), and trial-simulation engine (`trial.py`). PKComb-BOIN12 and TITE-PKComb-BOIN12 are additionally implemented in an interactive `R` Shiny application at https://22c7ba-damitri-kundu.shinyapps.io/PKComb-BOIN12-App/, supporting pre-trial calibration and real-time trial conduct without requiring local software installation.